\documentclass[twocolumn]{autart}    % Enable this line and disable the 
\usepackage{cite}
\usepackage{amsmath,amssymb,amsfonts}
\usepackage{algorithmic}
\usepackage{graphicx}
\usepackage{subcaption}
\usepackage{textcomp}
\usepackage[scr=rsfso,cal=zapfc,frak=euler,bb=ams]{mathalfa}
\usepackage{enumerate}

\usepackage{enumitem}
\usepackage{booktabs}
\usepackage{setspace}
\usepackage[english]{babel}
\usepackage[utf8]{inputenc}
\usepackage{fancyhdr}
\usepackage{yfonts}
\usepackage{mathrsfs}
\usepackage{xspace}
\usepackage{psfrag}
\usepackage{url}

\usepackage[dvipsnames]{xcolor}
\usepackage{xcolor}
\usepackage{lipsum}

\usepackage{verbatim}

\providecommand{\com[2]}{\begin{tt}[#1: #2]\end{tt}}
\usepackage{color}

\providecommand{\abs[1]}{\left|#1\right|}
\providecommand{\norm[1]}{\abs{\abs{#1}}}%

\newtheorem{theorem}{Theorem}
\newtheorem{lemma}[theorem]{Lemma}
\newtheorem{remark}{Remark}

\newtheorem{definition}{Definition}

\newtheorem{proposition}[theorem]{Proposition}

\DeclareSymbolFont{bbold}{U}{bbold}{m}{n}
\DeclareSymbolFontAlphabet{\mathbbold}{bbold}
\newcommand{\vect}[1]{\mathbbold{#1}}
\newcommand{\vectorones}[1][]{\vect{1}_{#1}}

\newcommand{\Med}{\textup{Med}}

\newcommand{\Ndiff}{N_{\textup{diff}}}
\newcommand{\Xeq}{X_{\textup{eq}}}
\newcommand{\maxv}[1]{\text{max}^{\,(#1)}}
\newcommand{\order}[1]{\text{argmax}^{\,(#1)}}
\newcommand{\C}{\mathcal{C}}

\newcommand{\M}{\mathcal{M}}
\newcommand{\N}{\mathcal{N}}

\newcommand{\G}{\mathcal{G}}
\newcommand{\V}{\mathcal{V}}

\newcommand{\X}{\mathcal{X}}

\newcommand{\argmin}{\operatorname{argmin}}
\newcommand{\argmax}{\operatorname{argmax}}

\graphicspath{{./fig/}}

\usepackage{graphicx}          % Include this line if your 
\begin{document}

\begin{frontmatter}

\title{Modeling, Analysis, and Control of Continuous-Time Weighted-Median Opinion Dynamics\thanksref{footnoteinfo}} % Title, preferably not more 

\thanks[footnoteinfo]{This manuscript is in draft status and remains to be further revised. This work is supported by the "Siddarta"  ARC at UCLouvain and by the Kornet FNRS research project.
}

\author[PKU]{Yi Han}\ead{han\_yi@stu.pku.edu.cn},    % Add the 
\author[UCL]{Julien M. Hendrickx}\ead{julien.hendrickx@uclouvain.be},               % e-mail address 
\author[CAS]{Ge Chen}\ead{chenge@amss.ac.cn}, 
\author[PKU]{Wenjun Mei\corauthref{cor1}}\ead{mei@pku.edu.cn}% (ead) as shown
%\author[ETH]{Florian D\"orfler}\ead{dorfler@ethz.ch},

\address[PKU]{Department of Control Science and Systems Engineering, Peking University}
\address[UCL]{ICTEAM institute, UCLouvain}  
\address[CAS]{Academy of Mathematics and Systems Science, Chinese Academy of Sciences} % (ead) as shown  % Please supply                                              
%\address[ETH]{Automatic Control Laboratory, ETH Zurich}            % full addresses    % here.
\corauth[cor1]{Corresponding author. Email: mei@pku.edu.cn}

% \title{The Continuous-Time Weighted-Median\\ Opinion Dynamics* 
% \thanks{*This manuscript is in draft status and remains to be further revised.}}
% \author{Yi Han, \IEEEmembership{Graduate Student Member, IEEE}, Wenjun Mei, \IEEEmembership{Member, IEEE}, Ge Chen, \IEEEmembership{Member, IEEE}, \\ Florian D\"orfler, \IEEEmembership{Member, IEEE}, Julien M. Hendrickx, \IEEEmembership{Member, IEEE}
% \thanks{Y. Han is with the Department of Mechanics and Engineering Science, Peking University (e-mail: han\_yi@stu.pku.edu.cn). W. Mei is with the Department of Mechanics and Engineering Science, Peking University (e-mail: mei@pku.edu.cn).G. Chen is with the Academy of Mathematics and Systems Science, Chinese Academy of Sciences (e-mail: chenge@amss.ac.cn). F. D\"orfler is with Automatic Control Laboratory, ETH Zurich (e-mail: dorfler@ethz.ch).  J. M. Hendrickx is with the ICTEAM institute, UCLouvain (e-mail: julien.hendrickx@uclouvain.be).}
% \thanks{ This work is supported by the "Siddarta"  ARC at UCLouvain and by the Kornet FNRS research project.} }

\begin{keyword}                           % Five to ten keywords,  
Social Networks; Opinion Dynamics; Weighted Median; Consensus.           % chosen from the IFAC 
\end{keyword}                             % keyword list or with the 
                                          % help of the Automatica 
                                          % keyword wizard
\begin{abstract}                          % Abstract of not more than 200 words.
Simple yet predictive mathematical models are essential for mechanistic understanding of opinion evolution in social groups. The weighted-median mechanism has recently been proposed as a well-founded alternative to conventional DeGroot-type opinion dynamics. However, the original weighted-median model excludes compromise behavior, as individuals directly adopt their neighbors’ opinions without forming intermediate values. In this paper, we introduce a parsimonious continuous-time extension of the weighted-median model by incorporating individual inertia, allowing opinions to move gradually toward the neighbors’ weighted median. Empirical evidence shows that this model outperforms both the original weighted-median and DeGroot models with inertia in predicting opinion shifts. We provide a complete theoretical analysis of the proposed dynamics: the equilibria are characterized and shown to be Lyapunov stable; global convergence is established via the Bony–Brezis method, yielding necessary and sufficient conditions for consensus from arbitrary initial states. In addition, we derive a graph-theoretic condition for persistent disagreement and a necessary and sufficient condition for steering the system to any prescribed consensus value through constant external inputs to a subset of individuals. These results reveal how a social group’s resilience to external manipulation fundamentally depends on its internal network structure.
\end{abstract}

\end{frontmatter}

\section{Introduction}
\subsection{Background and Motivations}

Opinion dynamics examine how social interactions shape public opinions. The French-DeGroot model, a cornerstone of the field, introduced weighted averaging into opinion updating\cite{JRPF:56,MHDG:74}. Numerous discrete- and continuous-time variants have since been analyzed\cite{AVP-RT:17}, yet the averaging mechanism has been criticized for its unrealistic assumption of strong attraction to distant opinions\cite{WM-FB-GC-JH-FD:22,BA-FA-LNE:23}. To address this issue, Mei et al.\cite{WM-FB-GC-JH-FD:22} proposed the weighted-median mechanism, which yields richer and more realistic dynamics. However, it excludes ``compromise'': individuals adopt neighbors’ opinions directly, preventing the emergence of intermediate values.

A natural remedy to the above problem is inertia, whereby individuals move toward rather than directly adopt the median. Motivated by this idea and supported by empirical data, we study a continuous-time weighted-median model. Statistical tests show that incorporating inertia improves predictive accuracy, and we provide a theoretical analysis of its equilibria, convergence, consensus conditions, and pinning controllability.

\subsection{Literature Review}

Opinion dynamics has been an important interdisciplinary research topic for about 70 years, drawing interest of researchers from economics\cite{QZ-GK-HZ:20}, mathematics\cite{MC-ACL-BP:15}, physics\cite{FG-EO-SB:19}, and control theory\cite{VA-FB-AKS:17}. For comprehensive surveys, see \cite{AVP-RT:17,AVP-RT:18,BDOA-MY:19}. One of the most celebrated mechanisms for opinion dynamics is the weighted-averaging rule introduced in the French-DeGroot model\cite{JRPF:56,MHDG:74}, where individuals repeatedly average their neighbors’ opinions. This model ensures consensus under mild network connectivity conditions\cite{FB:21} and has broad applications, e.g., in coordination algorithms \cite{WR-RWB:05,NA-LJ:17} and distributed optimization~\cite{AN-AO:14}. 

To account for persistent disagreement, several variants have been proposed. For examples, the Friedkin–Johnsen model introduces prejudice\cite{NEF-ECJ:90}; the Hegselmann–Krause model restricts interactions to bounded confidence intervals\cite{RH-UK:02,CB-CA-AP-FV:24}; and the Altafini model incorporates antagonistic relations\cite{CA:13}. Further extensions enrich the averaging framework. Multidimensional settings are studied in \cite{SEP-AVP-RT-NEF:17,NEF-AVP-RT-SEP:16}; nonlinear multioption dynamics with agent-option couplings are analyzed in \cite{BA-FA-LNE:23}; social learning is modeled by averaging Bayesian updates\cite{AJ-AS-ATS:10,DA-AO:11}; and the role of network structure is examined in \cite{FA-GD:04,PJ-AM-NEF-FB:13d,ML-HD:19}. 

Nevertheless, weighted averaging implies unrealistic attraction to distant opinions. To provide a fundamental resolution to this problem, a game-theoretic framework is proposed~\cite{WM-FB-GC-JH-FD:22}, which models opinion updates as best responses minimizing
\begin{align*}
    u_i(x_i) = \sum_{j=1 }^n w_{ij} \vert x_i - x_j\vert^\alpha,\quad \alpha>0.
\end{align*}
Here $\alpha$ tunes individuals' sensitivity to opinion distance: $\alpha=2$ recovers the classic DeGroot model, while $\alpha=1$ decouples attractiveness from distance, yielding the weighted-median mechanism\cite{WM-FB-GC-JH-FD:22}. This new mechanism is supported by experimental evidence and, in certain aspects, leads to more realistic macroscopic predictions than averaging-based models. A recent paper~\cite{WM-JMH-GC-FB-FD:24} rigorously analyzed this weighted-median model~\cite{WM-JMH-GC-FB-FD:24}, including its equilibria set, almost-sure convergence, and graph-theoretic conditions for reaching consensus and persistent disagreement. Extensions address prejudice\cite{RZ-ZL-GC-WM:24} and network-level approximations\cite{LM-PGH-MAP:24}, further establishing the weighted-median as a robust alternative micro-foundation for opinion dynamics.

\subsection{Contribution}

The main contributions of this paper are as follows:

Firstly, we propose the continuous-time weighted-median opinion dynamics model, extending the discrete formulation to continuous time while capturing compromise behavior. The model is also empirically supported by behavioral experiment data.

Secondly, we conduct a theoretical analysis. We prove existence and uniqueness of solutions by showing that the weighted-median operator is non-expanding, characterize all equilibria, and establish their stability. Using LaSalle’s invariance principle and a refinement of invariant sets, we prove convergence from any initial condition and derive necessary and sufficient conditions for consensus. We also provide a sufficient graph-theoretic condition under which almost all initial conditions converge to disagreement equilibria. Although our results parallel those of the discrete-time model~\cite{WM-JMH-GC-FB-FD:24}, the proof techniques differ fundamentally, as the former is a deterministic nonlinear ODE while the latter is a stochastic finite-state system with random asynchronous updates.

Thirdly, we study the problem of steering the weighted-median opinion dynamics to censensus by pinning the opinions of a subset of individuals. We show via theoretical analysis that cohesive sets function as the minimal units for control. By quantifying the minimal number of pinned individuals required for consensus, we analyze how controllability depends on network structure through simulations. The continuous-time weighted-median model displays more reasonable robustness to external manipulation than the weighted-averaging mechanism. Results show that denser or more randomized networks are significantly harder to control, in contrast with averaging-based models, where the system is fully manipulatable as long as the network is strongly connected and any single node is pinned.

\subsection{Organization}
After the introduction, Section II presents model setup and empirical support. Section III presents the theoretical analysis. Section IV addresses controllability issues via pinning control. Section V is the conclusion.

\section{Empirical Evidence and Model Setup} 

The original discrete-time weighted-median model does not capture compromise, as individuals directly adopt neighbors’ opinions. This limitation is resolved by continuous-time dynamics, where individuals move toward weighted-median opinions. This modification is supported by empirical evidence presented below.

\subsection{Empirical Support for the Weighted-Median Mechanism with Inertia}

\begin{figure}
\begin{center}
\includegraphics[width=1.0\linewidth]{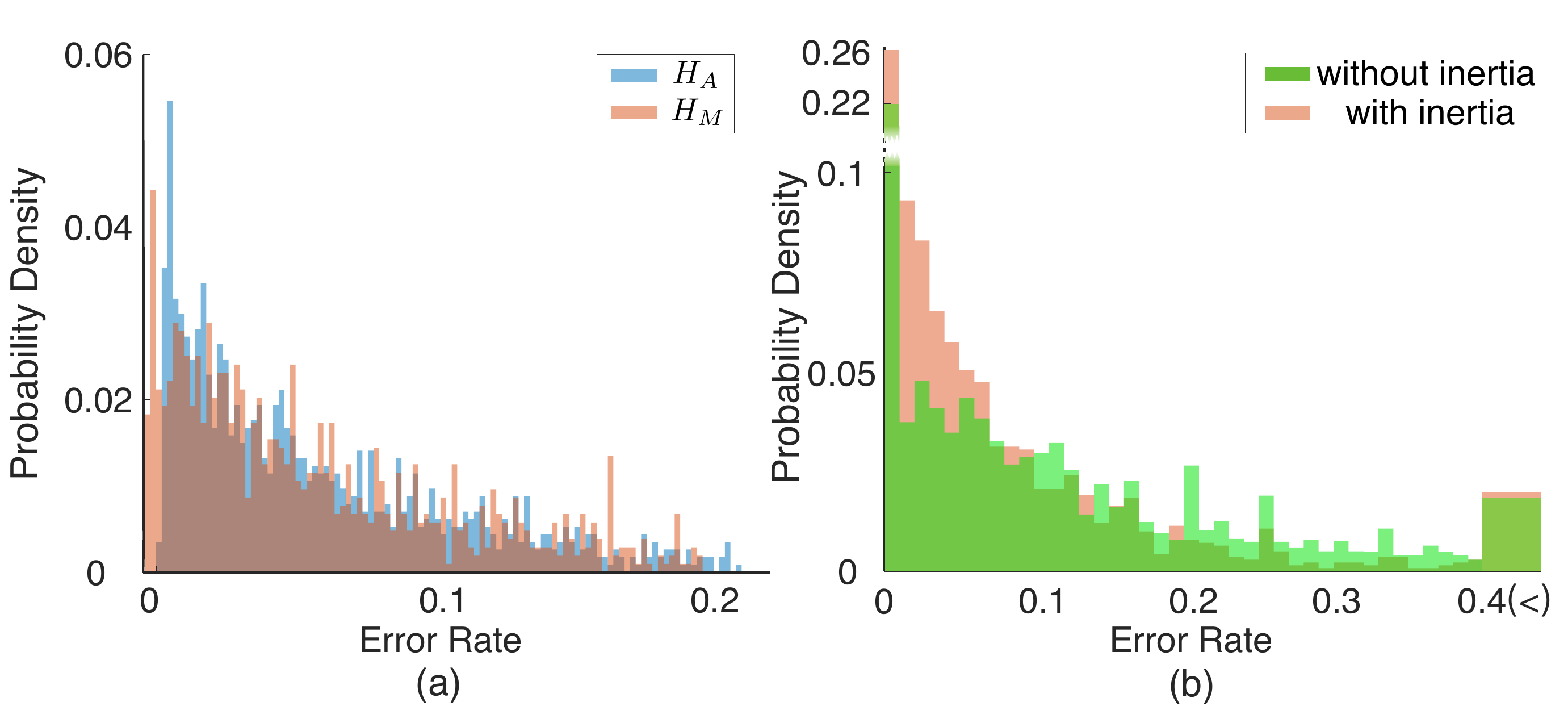}
\caption{%(a) Distributions of prediction error under $\text{H}_\text{A}$ and $\text{H}_\text{M}$ models after removing the top and bottom 10\% outliers for clearer visualization. (b)  Distributions of prediction error under weighted-based model      with inertia and weighted-based model without inertia after removing the top and bottom 10\% outliers for clearer visualization. A nonlinear axis scale is introduced to handle the extreme height of the first bin, which is visually truncated with a white gap. In addition, the last bin (0.4$<$) includes all error rate greater than 0.4 to account for tail values in a compact form.
(a) Distributions of prediction errors under the $\text{H}_\text{A}$ and $\text{H}_\text{M}$ models. (b) Distributions of prediction errors under the weighted-median model with and without inertia. The top and bottom 10\% outliers are not displayed in panel~(a) and~(b). In panel~(b), A nonlinear axis scale and a white gap are used to accommodate the first bin, which is disproportionaly high. The last bin (0.4$<$) aggregates all error rates greater than 0.4 to compactly represent the tail.
    }\label{fig:empirical-hist}
\end{center}
\end{figure}

We test our the weighted-median mechanism with inertia on human-subject data\cite{CVK-SM-PG-PJR-JMH-VDB:16}. The dataset includes 18 experiments, each with 5–6 participants and 30 questions. Participants first estimate a value, e.g., proportion of certain color in a picture, and submit their second-round estimations after seeing peers’ estimates. Then they submit the third-round estimations after seeing other participants’ second-round answers. We compare different update rules in predicting their estimation shifts. Let $x_i(t)$ be agent $i$’s estimate at round $t=1,2,3$, and consider: 
\begin{align*}
\text{H}_\text{M}: & x_i(t+1)=\gamma_i(t) x_i(t)+\left[1-\gamma_i(t)\right] \operatorname{Median}(x(t)), \\
\text{H}_\text{A}: & x_i(t+1)=\beta_i(t) x_i(t)+\left[1-\beta_i(t)\right] \text{Average}(x(t)),
\end{align*}
with inertia coefficients $\gamma_i(t),\beta_i(t)$ estimated via least squares on the first 20 questions. Predictions for the last 10 are then evaluated. 
As shown in Fig.~\ref{fig:empirical-hist}~(a), $\text{H}_\text{M}$ achieves lower mean error (0.1075) than $\text{H}_\text{A}$ (0.1121), a 4.1\% relative reduction, significant by a Wilcoxon test ($p < 10^{-7}$).We further compare the median model with and without inertia. For the last 10 questions, mean and median errors significantly drop from 0.1903 and 0.0833 to 0.1075 and 0.0400, respectively (Fig.~\ref{fig:empirical-hist}~(b)). These results confirm that inertia significantly improves prediction, motivating the continuous-time weighted-median opinion dynamics model.

\subsection{Basic Definitions And Model Setup}
Let $\subseteq$ ($\subset$ resp.) be the symbols for subset (proper subset resp.). Denote by $\mathbb{N}$ the set of natural numbers, i.e., $\mathbb{N}=\{0,1,2,\dots\}$. Let $\mathbb{Z}$ ($\mathbb{Z}_+$ resp.) be the set of (positive resp.) integers. Let $\mathbb{R}^n$ be the n-dimensional Euclidean space. Let $\vect{1}_n$ ($\vect{0}_n$ resp.) be the $n$-dimension vector whose entries are all ones (zeros resp.).

Consider a group of $n$ individuals labelled by $\V = \{1,\dots,n\}$. The interpersonal influence among these individuals are described by a directed and weighted graph $\G(W)$ associated with a row-stochastic matrix $W$, referred to as the \emph{influence matrix}. Here, $w_{ij} > 0$ means that node $i$ assigns $w_{ij}$ weight to node $j$. (In this paper, the terms node, agent, and individual are used interchangeably.) Denote by $\N_i$ the set of node $i$'s out-neighbours, i.e., $\N_i = \{j\in \mathcal{V}|w_{ij}\neq 0\}$, including $i$ itself if $w_{ii}\neq 0$. The formal definition of weighted median is given below.
\begin{definition}[Weighted median]\label{def:weighted-median-in-general}
Given any $n$-tuple of real values $x=(x_1,\dots, x_n)$ and any $n$-tuple of non-negative weights $w=(w_1,\dots, w_n)$ with $\sum_{i=1}^n w_i=1$, $x^*\in \{x_1,\dots, x_n\}$ is a weighted median of $x$ associated with the weights $w$ if $x^*$ satisfies 
   \begin{align*}
   \sum_{i:\, x_i<x^*} w_i \le 1/2,\quad \text{and} \quad \sum_{i:\, x_i>x^*}w_i \le 1/2.
   \end{align*}
\end{definition} 
We also say that $x^*$ is a weighted median of $x$ associated with $w$. The following lemma characterizes some basic properties on the uniqueness of the weighted median~\cite{WM-FB-GC-JH-FD:22}.
\begin{lemma}[Properties of weighted median]\label{lem:properties-weighted-median}
Given any $n$-tuple of real values $x=(x_1,\dots, x_n)$ and any $n$-tuple of non-negative weights $w=(w_1,\dots, w_n)$ with $\sum_{i=1}^n w_i=1$, 
\begin{enumerate}
\item \label{lem:properties-weighted-median:1} the weighted median of $x$ associated with $w$ is unique if and only if there exists $x^*\in \{x_1,\dots, x_n\}$ with
   \begin{align*}
   \sum_{i:\, x_i<x^*} w_i < \frac{1}{2},\quad \sum_{i:\, x_i=x^*} w_i >0,\quad \sum_{i:\, x_i>x^*}w_i <\frac{1}{2}.
   \end{align*} 
Such an $x^*$ is the unique weighted median;
\item \label{lem:properties-weighted-median:2} the weighted medians of $x$ associated with $w$ are not unique if and only if there exists $z\in \{x_1,\dots, x_n\}$ such that $\sum_{i:\, x_i<z} w_i = \sum_{i:\, x_i\ge z} w_i = 1/2$. In this case, $\underline{x}^*\in \{x_1,\dots, x_n\}$ is the smallest weighted median if and only if
\begin{align*}
 \sum_{i:\, x_i<\underline{x}^*} w_i < \frac{1}{2}, \quad \sum_{i:\, x_i=\underline{x}^*} w_i > 0,\quad \sum_{i:\, x_i>\underline{x}^*} w_i = \frac{1}{2},
\end{align*}
and $\overline{x}^*\in \{x_1,\dots, x_n\}$ is the largest weighted median if and only if 
\begin{align*}
\sum_{i:\, x_i<\overline{x}^*} w_i = \frac{1}{2}, \quad \sum_{i:\, x_i=\overline{x}^*} w_i > 0,\quad \sum_{i:\, x_i>\overline{x}^*} w_i < \frac{1}{2}.
\end{align*}
Moreover, for any $\hat{x}\in \{x_1,\dots, x_n\}$ with $\underline{x}^*<\hat{x}<\overline{x}^*$, $\hat{x}$ is also a weighted median and $\sum_{i: x_i = \hat{x}}w_i =0$.
\end{enumerate}
\end{lemma}

Motivated by the need to incorporate individuals' compromise behavior and supported by empirical evidence, we construct and study the following continuous-time weighted-median (CTWM) opinion dynamics.
\begin{definition}\label{def:CT-WM-OpDyn}
Given any directed and weighted influence network $\G(W)$ with $n$ individuals, denote by $x(t)\in \mathbb{R}^n$ the vector of the individuals' opinions at time $t$. Starting from any initial condition $x_0\in \mathbb{R}^n$, the continuous-time weighted-median (CTWM) opinion dynamics is given as
\begin{align}\label{eq:CT-WM-OpDyn}
\dot{x}(t) := f\big(x(t)\big) = \Med\big(x(t);W\big) - x(t),
\end{align} 
where $\Med_i(x(t);W)$ is the weighted median of $x(t)$ associated with the weights given by the $i$-th row of $W$, i.e., $(w_{i1},w_{i2},\dots, w_{in})$. If the weighted-median is not unique, then we let $\Med_i\big(x(t);W\big)$ be the weighted median that is the closest to $x_i(t)$.
\end{definition}

Since the model proposed above is an ODE system instead of a finite-state machine, the proof methods used in the original weighted-median model, i.e., establishing convergence by manually designing an update sequence for any initial condition, is no longer applicable. In this paper, we analyze the asymptotic behavior of the continuous-time model via invariant-set arguments, which are completely different from what is used to analyze the original model.

Frequently used notations are introduced in Table~\ref{table:notations} and illustrated via the following example: Given a five-tuple 
\begin{align*}
    x=(0.7,-0.8,0.4,-0.5,0.4,-0.5,0)
\end{align*}
with 5 distinct values, $\text{Ndiff} (x)=5$ and the largest value is $\text{max}^{(1)}(x)=0.7$, which is achieved by the index $\text{argmax}^{(1)}(x)=\{1\}$. The smallest value is $\text{max}^{(5)}(x)=-0.8$ and is achieved by the index $\text{argmax}^{(5)}(x)=\{2\}$. The indices corresponding to the three largest (distinct) values are $\text{argmax}^{(\leq 3)}(x)=\text{argmax}^{(< 4)}(x)=\{1,3,5,7\}$. The indices corresponding to the three smallest (distinct) values are $\text{argmax}^{(\geq 3)}(x)=\text{argmax}^{(> 2)}(x)=\{2,4,6,7\}$.

\begin{table}[htbp]\caption{Notations frequently used in this paper}\label{table:notations}
\begin{center}
\begin{tabular}{r p{5.2cm} }
\toprule
$\phi(t,x_0)$ & the trajectory along the dynamics~\eqref{eq:CT-WM-OpDyn} starting from $x_0$ at time 0.\\ 
$\partial B$ & the boundary of the set $B$.\\
$\bar \M$ & $\M$'s complement, i.e. $\bar{\M}=\V\setminus \M$.\\
$\Ndiff(x)$ &  the number of distinct values among the entries of $x$. \\
$\maxv{k}(x)$ &  the $k$-th largest value among the $\Ndiff(x)$ distinct values of the entries of $x$ ($k\in \{1,\dots, \Ndiff(x)\}$). \\
$\order{k}(x)$ &  indices set $\{i\,|\, x_i = \maxv{k}(x)\}$, e.g., $\order{1}(x) = \argmax_k x_k$, $\order{\Ndiff(x)}(x) = \argmin_k x_k$. Let $\order{0}(x)$ be empty.\\
$\order{\le r}(x)$  &  $\cup_{s\le r}\, \order{s}(x)$. \\
$\order{\ge r}(x)$ &  $\cup_{s\ge r}\, \order{s}(x)$. ($\order{\ge \Ndiff(x)+1}(x)$ is empty.)\\
$\order{< r}(x)$  &  $\cup_{s< r}\, \order{s}(x)$. \\
$\order{> r}(x)$ &  $\cup_{s> r}\, \order{s}(x)$.\\ 
\bottomrule
\end{tabular}
\end{center}
\end{table}

\section{Dynamical behavior of the continuous-time weighted-median opinion dynamics}

In this section, we establish the existence and uniqueness of system~\eqref{eq:CT-WM-OpDyn}'s solution and characterize the equilibria set. A necessary and sufficient graph-theoretic condition for reaching consensus and a sufficient condition for convergence to disagreement equilibria are also provided. 

Similar to the original discrete-time weighted-median model~\cite{WM-JMH-GC-FB-FD:24}, the behavior of the CTWM opinion dynamics depends also on the following two network structures: cohesive sets and decisive links, reviewed below.

\begin{definition}[Cohesive set]\label{def:CS-MCS}
Given an influence network $\mathcal{G}(W)$ with node set $V$, a cohesive set $\mathcal{M}\subset \mathcal{V}$ is a subset of nodes that satisfies $\sum_{j\in \mathcal{M}} w_{ij}\ge 1/2$ for any $i\in \mathcal{M}$. A cohesive set $\M$ is a maximal cohesive set if there does not exist $i\in \V\setminus \M$ such that $\sum_{j\in \M} w_{ij}>1/2$.
\end{definition}

\begin{definition}[Decisive link and decisive graph]\label{def:decisive-indecisive-links}
Given an influence network $\mathcal{G}(W)$ with the node set $\V$, define the out-neighbor set of each node $i$ as $\N_i = \{j\in \V\,|\, w_{ij}\neq 0\}$. A link $(i,j)$ is a decisive out-link of node $i$, if there exists a subset $\theta\subset \N_i$ such that the following three conditions hold: (1) $j\in \theta$; (2) $\sum_{k\in \theta} w_{ik} > 1/2$; (3) $\sum_{k\in \theta\setminus \{j\}} w_{ik}<1/2$. Otherwise, the link $(i,j)$ is an indecisive out-link of node $i$. The \emph{decisive graph}
$\mathcal{G}_{\text{decisive}}(W)$ is the subgraph obtained by
removing all indecisive links.
\end{definition}

Basically, a cohesive set is a clique of nodes that will keep their opinions unchanged whenever consensus is reached within the clique. Decisive links are the links that could be the "tie-breaker" in determining a node's weighted-median opinion. The following lemma presents some important properties of cohesive sets. For the proof of Lemma~\ref{lem:properties-cohesive-expansion} and more properties, we refer to \cite{WM-JMH-GC-FB-FD:24}.
\begin{lemma}[{\cite[Lemma~4 \& Lemma~5]{WM-JMH-GC-FB-FD:24}}]\label{lem:properties-cohesive-expansion}
Given an influence network $\mathcal{G}(W)$ with the node set $\V$, the following statements hold:
\begin{enumerate}
\item If $\M_1,\,\M_2\subseteq \V$ are both cohesive sets, then $\M_1\cup \M_2$ is also a cohesive set;
\item if $\M$ is maximally cohesive and $\M\subset \V$, then $\V\setminus \M$ is also maximally cohesive.
\end{enumerate}
\end{lemma}

\subsection{Lemmas on Positive Invariance}

The proofs of the main results in this paper mainly rely on the analysis of positive invariant sets and limit points, involving the following useful concepts and lemmas.

For any autonomous system $\dot{x}=f(x)$, denote the $\omega$-limit set of any $x$ by $\omega(x)$.   A positively invariant set for the autonomous system is a set  $S$  such that, if  $\phi(0,x) \in S$,  then the solution  $\phi(t,x)\in S$  for all  $t \geq 0$.
The following lemma states that, if the trajectory of an ODE system converges to a set of stable equilibria, then it converges to a unique equilibrium. Its proof is given in Appendix~\ref{append:proof-lem:CT-from-eq-set-to-single-eq}.
\begin{lemma}[Convergence to unique equilibrium]\label{lem:CT-from-eq-set-to-single-eq}
Consider an ODE system $\dot{x}=f(x)$ with domain $D\subseteq \mathbb{R}^n$. Let $\chi$ be a set of stable equilibria of the system. Given any $x_0\in D$, if its $\omega$-limit set satisfies $\omega(x_0)\subseteq \chi$, then the solution trajectory $\phi(t,x_0)$ converges to a unique $x^*\in \chi$, i.e., $\omega(x_0)=\{x^*\}$.
\end{lemma}

We recall a classical result by Bony and Brezis~\cite{blanchini1999set,redheffer1972theorems}, which characterizes positively invariant sets by examining the vector field on the boundary. The key idea is that trajectories cannot exit a set if the vector field always points inward or tangential to the boundary. The following lemma provides the formal statement.

\begin{lemma}[Bony-Brezis theorem]\label{thm:Bony-Brezis}
Let $C\subset \mathbb{R}^n$ be a closed subset and let $F:\mathbb{R}^n\to \mathbb{R}^n$ be a Lipschitz continuous vector field. Then $C$ is positively invariant for the system $\dot x = F(x)$ if and only if $F(x)^{\top} v\leq 0$ for any $x\in \partial C$ and any exterior normal vector $v$ at $x$, i.e., any $v\in \mathbb{R}^n$ such that $v^{\top} (y - x) \leq 0$ holds for all $y \in C$.
\end{lemma}
Lemma~\ref{thm:Bony-Brezis} provides a general condition for the positive invariance of a closed set under a Lipschitz continuous vector field. Building upon this result, the following proposition establishes a specific case where a convex set remains positively invariant under system $\dot x = F(x) - x$.

\begin{proposition}\label{prop:fwd_inv_generic_short}
Let $X \subset \mathbb{R}^n$ be a closed convex set and $F=\mathbb{R}^n\to\mathbb{R}^n$ a Lipschitz continuous vector field with the property that for all $x\in X, F(x)\in X$. Then the set $X$ is positively invariant for the system $\dot x = F(x) - x$.
\end{proposition}
\emph{Proof:} Since $X$ is convex and $F(X)\subseteq X$, the segment from $x$ to $F(x)$ is included in $X$ for every $x$. Hence by convexity, $(F(x)-x)^{\top}v\leq 0$ for any exterior normal vector $v$ at $x$, and the result follows directly from Lemma~\ref{thm:Bony-Brezis}. 
\hfill\qed

The lemma below helps to further refine the set that the solution converges to. The proof is given in Appendix~\ref{append:proof-lem:refine-omega-limit-set}.
\begin{lemma}\label{lem:refine-omega-limit-set}
Consider the system $\dot{x}=f(x)$, where $f:\mathbb{R}^n\to \mathbb{R}^n$ is globally Lipschitz and $G\subseteq \mathbb{R}^n$ is compact and positively invariant with respect to the system dynamics. For any given $x\in G$, if there exists an open set $B\subseteq G$ such that: 1) $x\notin B\cup \partial B$; 2) there exists $T>0$ such that $\phi(T,x)\in B$; 3) $B\cup \partial B$ is positively invariant, then $x\notin \omega(y)$ for any $y\in G$.
\end{lemma}

\begin{lemma}\label{lem:max-is-glob-Lipschitz}
For any $x\in \mathbb{R}^n$, the function $V:\, \mathbb{R}^n\to\mathbb{R}$ defined by $V(x)=\max\,\{x_1,\dots,x_n\}$ is globally Lipschitz.
\end{lemma}

\emph{Proof:}  For any $x,\, y\in \mathbb{R}^n$, $|V(x)-V(y)|=|\max_i x_i - \max_j y_j|$. Let $i^*\in \argmax_i x_i$ and $j^*\in \argmax_j y_j$ and, without loss of generality, assume that $\max_i x_i \ge \max_j y_j$. We have
\begin{align*}
0 & \le |V(x)-V(y)| = |x_{i^*}-y_{j^*}|=x_{i^*}-y_{j^*}\le x_{i^*}-y_{i^*} \\
& \le \lVert x-y \rVert_{\infty}.
\end{align*}
Therefore, $V(x)$ is globally Lipschitz in $x$.
\hfill\qed

\subsection{Main results}

We first establish the existence and uniqueness of the solution to the continuous-time weighted-median opinion dynamics~\eqref{eq:CT-WM-OpDyn} starting from any initial condition.
\begin{theorem}[Existence and uniqueness of solution]\label{thm:Med-NonExpanding}
Given the influence matrix $W$, $\Med(x;W)$ is non-expanding in the sense that, for any $x$, $y\in \mathbb{R}^n$, $\lVert \Med(x;W)-\Med(y;W) \rVert_{\infty} \le \lVert x-y \rVert_{\infty}$. Therefore, the function $f$ in equation~\eqref{eq:CT-WM-OpDyn} is globally Lipschitz in $\mathbb{R}^n$. As a result, for any initial condition $x_0\in \mathbb{R}^n$, the solution $\phi(t,x_0)$ exists and is unique.
\end{theorem}
\emph{Proof of Theorem~\ref{thm:Med-NonExpanding}:}
We prove by contradiction. Suppose that $\Med(x;W)$ is not non-expanding, i.e. there exist $x$, $y\in \mathbb{R}^n$ such that 
\begin{align*}
\lVert \Med(x;W)-\Med(y;W) \rVert_{\infty} > \lVert x-y \rVert_{\infty}.
\end{align*}
As a consequence, there exists some $i\in \mathcal{V}$ such that
\begin{align*}
|\Med_i(x;W)-\Med_i(y;W)|>\lVert x-y \rVert_{\infty}.
\end{align*}
Without loss of generality, assume that $\Med_i(x;W)>\Med_i(y;W)$. Let 
\begin{align*}
\mathcal{X}^\geq & = \big{\{} j\in \mathcal{N}_i\,\big|\,x_j\ge \Med_i(x;W) \big{\}},\\
\mathcal{Y}^\leq&=\big{\{} j\in \mathcal{N}_i \,\big|\,y_j\le \Med_i(y;W) \big{\}}.
\end{align*}
According to Definition~\ref{def:weighted-median-in-general},
\begin{align*}
\sum_{j\in \mathcal{X}^\geq} w_{ij}\ge 1/2,\quad \text{and }\sum_{j\in \mathcal{Y}^\leq}w_{ij}\ge 1/2.
\end{align*}
Now we show that the two inequalities above lead to contradiction.

We first consider the case when the strict inequality holds in at least one of the two inequalities above. Since $\sum_{j\in \mathcal{N}_i} w_{ij}=1$ and $w_{ij}\ge 0$ for any $j$, we have that $\mathcal{X}^\geq\cap \mathcal{Y}^\leq$ must be non-empty, otherwise we obtain
\begin{align*}
1<\sum_{j\in \mathcal{X}^\geq \cup \mathcal{Y}^\leq}w_{ij}\le \sum_{j=1}^n w_{ij},
\end{align*} 
which cannot be true. For any $j^*\in \mathcal{X}^\geq\cap \mathcal{Y}^\leq$, since $x_j\ge \Med_i(x;W)$ and $y_j\le \Med_i(y;W)$, we have 
\begin{align*}
x_{j^*}-y_{j^*}\ge \Med_i(x;W)-\Med_i(y;W)>0,
\end{align*}
which contradicts the assumption that $\big| \Med_i(x;W)-\Med_i(y;W) \big|> |x_k-y_k|$ for any $k\in \mathcal{V}$. 

Now we consider the case when $\sum_{j\in \mathcal{X}^\geq}w_{ij}=1/2$ and $\sum_{j\in \mathcal{Y}^\leq}w_{ij}=1/2$. 
Let us first prove that $x_i\ge \Med_i(x;W)$. Since 
\begin{align*}
\sum_{j\in \mathcal{X}^\geq}w_{ij} = \sum_{j:\, x_j\ge \Med_i(x;W) }w_{ij}=\frac{1}{2},
\end{align*}
it follows from \ref{lem:properties-weighted-median:2}) of Lemma~\ref{lem:properties-weighted-median} that the weighted median of $x$ associated with the $i$-th row of $W$ is not unique, and $\Med_i(x;W)$ is not the smallest among all such weighted medians. Hence $\Med_i(x;W)$ is either the largest weighted median or a weighted median between the largest and the smallest. According to Definition~\ref{def:weighted-median-in-general}, if $\Med_i(x;W)$ is the largest, then $x_i\ge \Med_i(x;W)$; if $\Med_i(x;W)$ is between the largest and the smallest, then $x_i = \Med_i(x;W)$. In both cases, we have $x_i\ge \Med_i(x;W)$. A parallel argument shows that $y_i \le \Med_i(y;W)$. Consequently, we have that 
\begin{align*}
x_i-y_i \ge \Med_i(x;W)-\Med_i(y;W),
\end{align*}
which yields a contradiction to our assumption.

As both cases result in a contradiction, the assumption must be false. Hence, the map \(\Med(s;W)\) is non-expanding. As straightforward results, the function $f(x)=\Med(x;W)-x$ is globally Lipschitz in $\mathbb{R}^n$ and the solution to the dynamics~\eqref{eq:CT-WM-OpDyn} exists and is unique.\qed

The following theorem characterizes the set of all the equilibria and their stability.
\begin{theorem}[Equilibria and Stability]\label{thm:CT-EqSet}
$\!$Consider the CTWM opinion dynamics~\eqref{eq:CT-WM-OpDyn} with the influence matrix $W$. The set of all the equilibria is given by the set:
\begin{align*}
 \Xeq = \Big{\{} & x\in \mathbb{R}^n \,\Big|\, \order{\le r}(x)\text{ is a maximal cohesive}\\
&\text{set on }\mathcal{G}(W),\text{ for any }r\in \big{\{}1,\dots,\Ndiff(x)\big{\}} \Big{\}}
\end{align*} 
Moreover, each equilibrium $x\in \Xeq$ is stable. 
\end{theorem}
 
Since the CTWM opinion dynamics and the original weighted-median dynamics have the same equilibria set, which is fully characterized by Theorem~6 in \cite{WM-JMH-GC-FB-FD:24}, the proof of Theorem~\ref{thm:CT-EqSet} in this paper focuses only on the stability of equilibria, based on the following proposition.

\begin{proposition}[Invariance for cohesive sets]\label{prop:positive_invar_cohesive}
Consider the weighted-median opinion dynamics~\eqref{eq:CT-WM-OpDyn} with the influence matrix $W$. If $\C$ is a cohesive set on the influence network $\mathcal{G}(W)$, then, for any $a,b\in \mathbb{R}\cup\{\pm \infty\}$, the set 
$$
X_{\C,[a,b]}:=\{x\in \mathbb{R}^n: x_i\in [a,b],\forall i\in \C\}
$$
is positively invariant. In particular, for any $x\in \mathbb{R}^n$, $\max_{k\in \C}\phi_k(t,x)$ is non-increasing and $\min_{k\in \C} \phi_k(t,x)$ is non-decreasing, for any $t\ge 0$.
\end{proposition}

The proof of Proposition~\ref{prop:positive_invar_cohesive} is provided in Appendix~\ref{append:proof-prop:positive_invar_cohesive}. As a consequence, if $\C$ is a cohesive set in $\G(W)$, then, for any $x\in \mathbb{R}^n$, the components of $\phi(t, x)$ corresponding to $\C$ are bounded in the compact set $[\min_{k\in \C}x_k(t), \, \max_{k\in \C} x_k(t)]^{|\C|}$ for any $t\ge 0$. With Proposition~\ref{prop:positive_invar_cohesive}, Theorem~\ref{thm:CT-EqSet} is proved as follows.

\emph{Proof of Theorem~\ref{thm:CT-EqSet}:} 

Recall that Theorem~6 in \cite{WM-JMH-GC-FB-FD:24} establishes that 
$x^* = \Med(x^*;W)$ if and only if 
$x^*$ is either a consensus vector 
or satisfies the following condition: for any $y$ such that 
$\min_i x^*_i < y \le \max_i x^*_i$, both 
$\{i \in \V \mid x^*_i < y\}$ and $\{i \in \V \mid x^*_i \ge y\}$ 
are maximal cohesive sets on $\G(W)$.  By Definition~\ref{def:CS-MCS} and 2) of Lemma~\ref{lem:properties-weighted-median}, this characterization 
is equivalent to the condition that for all $r \in \{1,\dots,\Ndiff(x^*)\}$, $\order{\le r}(x^*)$ is a maximal cohesive set on $\G(W)$.
Thus we have the equilibria set $\Xeq$ and now proceed to show their stability. Let $x^*\in \Xeq$. We will show that for every $\epsilon> 0$ and $x_0$ s.t. $\norm{x_0-x^*}_\infty\leq \epsilon$, there holds $\norm{\phi(t,x_0)-x^*}_\infty\leq \epsilon$ for all $t\geq 0$.

   Consider an arbitrary index $i\in \V$ and suppose $x_i^*\in \order{r}(x^*)$. By the definition, $i$ belongs to both $\order{ \leq r}(x^*)$ and $\V \setminus \order{ < r}(x^*)$, which are both maximal cohesive sets, the former by definition of $\Xeq$ and the latter by Lemma \ref{lem:properties-cohesive-expansion} applied to $\order{ < r}(x^*)= \order{ \leq r-1}(x^*)$.  Moreover, $x_i^*\leq x_j^*$ for every $j\in \order{ \leq r}(x^*)$ and $x_i^*\geq  x_j^*$ for every $j\in \V \setminus \order{ < r}(x^*)$.

    By the construction of $x_0$ that $\norm{\phi(t,x_0)-x^*}_\infty\leq \epsilon$ and the definition of $\order{ \leq r}(x^*)$, there holds $$x_{0,j}\geq x_j^*-\epsilon \geq x_i^*-\epsilon ,\quad \forall j \in \order{ \leq r}(x^*).$$ Since $\order{ \leq r}(x^*)$ is a cohesive set, Proposition \ref{prop:positive_invar_cohesive} implies that for all $j\in \order{ \leq r}(x^*) $, $$\phi_j(t,x_0)\geq x_i^*-\epsilon ,\quad \forall t\geq 0,$$ and in particular for $j=i$. Applying exactly the same argument to $\V \setminus \order{ < r}(x^*)$ shows that $$\phi_{i}(t,x_0)\leq x_i^*+\epsilon,\quad \forall t\geq 0.$$
    
    Combining the lower and upper bounds, we conclude that
$$x_i^* - \epsilon \leq \phi_i(t, x_0) \leq x_i^* + \epsilon, \quad \forall t \geq 0.$$
Since $i$ was chosen arbitrarily, this holds for all $i \in \V$, and hence $\|\phi(t, x_0) - x^*\|_\infty \leq \epsilon,\quad \forall t \geq 0.$
This establishes the claim.  
\hfill \qed

In what follows, we establish the convergence of the continuous-time weighted-median opinion dynamics. The proof is presented in Appendix~\ref{append:proof-thm:CT-convergence}.
\begin{theorem}[Convergence]\label{thm:CT-convergence}
Consider the weighted-median opinion dynamics defined by equation~\eqref{eq:CT-WM-OpDyn} with the influence matrix $W$. For any $x_0\in \mathbb{R}^n$, there exists $x^*\in \Xeq$, depending on $x_0$, such that the solution $\phi(t,x_0)$ converges to $x^*$ as $t\to \infty$.
\end{theorem}

The following theorem presents a necessary and sufficient graph-theoretic condition for converging to consensus equilibria, together with a sufficient condition for convergence to disagreement equilibria. 
\begin{theorem}[Consensus and disagreement]\label{thm:CT-consensus-disagreement}
Consider the weighted-median opinion dynamics defined by equation~\eqref{eq:CT-WM-OpDyn} with the influence matrix $W$. The following statements hold:
\begin{enumerate}
\itemsep0em

\item $\phi(t,x_0)$ converges to a consensus equilibrium for all initial condition $x_0\in \mathbb{R}^n$ if and only if $\mathcal{V}$ is the only maximal cohesive set on $\mathcal{G}(W)$.
\item If $\mathcal{G}_{\text{decisive}}(W)$, as defined in Definition~\ref{def:decisive-indecisive-links}, does not have a globally reachable node, then for almost every $x_0\in \mathbb{R}^n$ the solution $\phi(t,x_0)$ converges to a disagreement equilibrium.
%\item \textcolor{red}{conjecture: }If $\mathcal{G}_{\text{decisive}}(W)$ has a globally reachable node, then there exists a subset $X_{0,\text{consensus}}\subseteq \mathbb{R}^n$ with non-zero measure such that, for any $x_0\in X_{0,\text{consensus}}$, the solution $\phi(t,x_0)$ converges to a consensus equilibrium.
\end{enumerate}
\end{theorem}

To prove statement 2) of Theorem~\ref{thm:CT-consensus-disagreement}, Proposition~\ref{prop:continuity-measurability} is introduced. Proposition \ref{prop:continuity-measurability} shows the mapping from initial conditions to the equilibrium is continuous and measurable. The proof is provided in Appendix \ref{append:proof-prop:continuity-measurability}.

\begin{proposition}\label{prop:continuity-measurability}

Consider the ODE $\dot x(t) = f(x(t))$ for a globally Lipschitz continuous function $f:\mathbb{R}^n\to \mathbb{R}$, let $\phi(t,x)$ be the value taken at time $t$ by a solution to that ODE with initial condition $x\in \mathbb{R}^n$, and suppose that the following two conditions hold
\begin{itemize}
   \item The system converges, i.e. for every $x$, $\bar \phi(x) = \lim_{t\to \infty} \phi(t,x)$ is well defined. 
    \item Every equilibrium is stable: if $f(x^*)  = 0$ then for every $\epsilon>0$  there exists a $\delta>0$ such that for every $x\in B(x^*,\delta)$, $\phi(t,x)\in B(x^*,\epsilon)$ for all $t>0$.
    \end{itemize}
Then $\bar\phi $ is continuous, and therefore measurable.
\end{proposition}

\emph{Proof of Theorem~\ref{thm:CT-consensus-disagreement}:} (1) We first prove that if $\mathcal{V}$ is the only maximal cohesive set on $\mathcal{G}(W)$, then for any $x_0\in \mathbb{R}^n$, $\phi(t,x_0)$ converges to a consensus equilibrium. By Theorem~\ref{thm:CT-convergence}, for any $x_0 \in \mathbb{R}^n$, the trajectory $\phi(t, x_0)$ converges to some $x^* \in \Xeq$. Moreover, Theorem~\ref{thm:CT-EqSet} shows that such an equilibrium $x^*$ satisfies the property that $\order{1}(x^*)$ is a maximal cohesive set of $\mathcal{G}(W)$. Since $\mathcal{V}$ is the unique maximal cohesive set, it follows that $\order{1}(x^*) = \mathcal{V}$, which implies that $x^*$ is a consensus equilibrium.
Next, we show that if $\mathcal{V}$ is not the only maximal cohesive set on $\mathcal{G}(W)$, then there exists a subset $X_{0,\text{disagree}}\subseteq \mathbb{R}^n$ with non-zero measure, such that, for any $x_0\in X_{0,\text{disagree}}$, $\phi(t,x_0)$ converges to a disagreement equilibrium. Suppose $\mathcal{V}_1\subset \mathcal{V}$ is a maximal cohesive set, then its complement $\mathcal{V}_2=\mathcal{V}\setminus \mathcal{V}_1$ is also maximally cohesive. The set
\begin{align*}
\X = \big\{ x\in \mathbb{R}^n \,\big|\, x_i>x_j\text{ for any }i\in \mathcal{V}_1,\, j\in \mathcal{V}_2 \big\}
\end{align*}
is a subset of $\mathbb{R}^n$ with non-zero measure. For any such $x\in \X$, according to Lemma~\ref{lem:properties-cohesive-expansion}, 
\begin{align*}
\min_{k\in \mathcal{V}_1} \phi_k(t,x) \ge \min_{k\in \mathcal{V}_1} x_k > \max_{k\in \mathcal{V}_2} x_k \ge \max_{k\in \mathcal{V}_2} \phi_k(t,x).
\end{align*}
Therefore, $\phi(t,x)$ does not converges to any consensus equilibrium for all $x\in \X$. This concludes the proof of statement~1).

(2) As stated in \cite{WM-JMH-GC-FB-FD:24}, for a node $j$ to be influenced by the opinion $x_i(t)$ for all $t\geq 0$, there must exist at least one path from $j$ to $i$ on the decisive influence network $\mathcal{G}_{\text{decisive}}(W)$. The digraph $\mathcal{G}_{\text{decisive}}(W)$ has no globally reachable node if and only if there exist at least two disjoint groups of nodes such that no node in one group has an outgoing edge to any node in the other group\cite{ZL-FB-MM:05}. Thus, there exist two disjoint subsets of nodes, denoted by $S_1,S_2$, which results in two independent (completely decoupled) dynamical subsystems. Although additional nodes may exist outside $S_1 \cup S_2$, we first restrict attention to the case $S_1\cup S_2=\V$, as the extension to general cases is straightforward.  

Let $X_1 \in \mathbb{R}^{n_1},X_2\in \mathbb{R}^{n_2}$ represent the initial conditions of the nodes in $S_1$ and $S_2$, respectively. Since the dynamics restricted to $S_1$ form an internally self-sustained subsystem, the trajectories of its nodes depend only on $X_1$, denoted by $\phi^1(t, X_1)$, with limiting value $\bar{\phi^1}(X_1) = \lim_{t \to \infty} \phi^1(t, X_1)$. An analogous construction applies to $S_2$. Based on these limits, we define the mapping $\psi:\mathbb{R}^{n_1}\times \mathbb{R}^{n_2}:(X_1,X_2)\to R^2:\psi(X_1,X_2) = (\psi_1(X_1),\psi_2(X_2)) $, with  $$\psi_1(X_1)\!:=\!\frac{1}{|S_1|}\sum_{i=1}^{n_1}\bar{\phi_i^1}(X_1) ,\qquad
\psi_2(X_2)\!:=\!\frac{1}{|S_2|}\sum_{i=1}^{n_2}\bar{\phi_i^2}(X_2).$$ Moroever, it is continuous by Proposition \ref{prop:continuity-measurability} and linearity of averaging.

If global consensus were to emerge, then necessarily $\psi_1(X_1)=\psi_2(X_2)$, i.e., the group averages of the limiting values must coincide. We therefore focus on the set of initial conditions satisfying this necessary condition: $$M:=\psi^{-1}( span \{(1,1)\})$$ and show that $M$ has measure 0. Since $\psi$ is continuous and $span \{(1,1)\}$ is a Borel set, then $M$ is a Borel set and therefore Lebesgue measurable. 

Next, for any $\alpha \in \mathbb{R}$, define
$$
M_\alpha = \{(X_1+\alpha\vect{1}_{n_1},X_2): (X_1,X_2) \in M\}.
$$
$M_\alpha$ is the set of initial conditions obtained by translating, from an element of $M$, the value of all agents in $S_1$ by a same constant $\alpha$. By translational invariance of measure, we have $\mu(M_\alpha) = \mu(M)$,where $\mu$ denotes the Lebesgue measure. Moreover, since the dynamics in $S_1$ and $S_2$ are independent and invariant under translation, a shift by $\alpha$ in $S_1$ will shift the limiting value by $\alpha$ for $S_1$. In other words, for any $(X_1,X_2)\in M_\alpha$, we have $\psi_1(X_1)-\psi_2(X_2) = \alpha$. This implies that all the $M_\alpha$ are disjoint for all $\alpha \in \mathbb{R}$. 

Finally, since all the $M_\alpha$ ($\alpha \in \mathbb{R}$) are disjoint and have the same measure $\mu(M)$, we must have $\mu (M) =0$; otherwise $\mu(M) > 0$, then the sum $\sum_{\alpha \in \mathbb{R}} \mu(M_\alpha)$ diverges because there are uncountably many disjoint sets $M_\alpha$, contradicting the finite measure of any bounded subset of $\mathbb{R}^{n_1} \times \mathbb{R}^{n_2}$. 

The fact $\mu (M) =0$ implies that trajectories originating from almost every initial condition convergence to disagreement. Moreover, in the presence of additional nodes outside $S_1 \cup S_2$, the dimension of the system increases and consensus requires even stricter alignment across groups. This only further reduces the measure of consensus-inducing initial conditions. Consequently, the consensus set remains of measure zero in the general case. This concludes the proof of statement~2).\hfill \qed

\begin{remark}
The condition that $\mathcal{V}$ is not the only maximal cohesive set on $\mathcal{G}(W)$ does not preclude the existence of a non-zero measure subset of initial conditions starting from which the trajectory converges to consensus. For example, consider a network with three nodes and the weighted adjacency matrix $W$ defined as
\[
W = \begin{bmatrix} 
0.4 & 0.2 & 0.4 \\ 
0 & 1 & 0 \\ 
0.3 & 0.3 & 0.4 
\end{bmatrix}
\]  Here, the set $\{1,3\}$ forms another maximal cohesive set. When the initial opinion $x^0$ satisfies $x^0_1\leq x^0_2\leq x^0_3$, then the weighted-median update rule yields $Med(W;x(t))\equiv x^0_2 \vect{1}_3$, which implies that the system converges to the consensus equilibrium, i.e. $\lim_{t \to \infty} x(t)=x^0_2 \vect{1}_3$. That is, while disagreement equilibria are possible, consensus may still be attainable from certain initial conditions. 
\end{remark}

\section{Further Discussion and Simulations}

A key difference between the median-based opinion dynamics and the averaging-based opinion dynamics is that the former exhibit more reasonable robustness to external manipulation. Consider a pinning-control scenario where a group of $n$ individuals exchange opinions over an influence network $\G(W)$, and a subset of them is manipulated such that they are persistently influenced by a fixed external opinion input. Let $\V_p$ denote the set of manipulated individuals and $\V_u$ the set of unmanipulated ones. The opinion dynamics can be described as
\begin{equation}\label{eq:prejudiced-dynamic}
        \dot{x}_i(t) = \gamma_i x^* + (1-\gamma_i)\Big( f_i(x(t);W)-x_i(t)\Big),\,\, i\in \V, 
\end{equation}
where $\gamma_i\in (0,1]$ for any $i\in \V_p$ and $\gamma_i=0$ for any $i\in \V_u$. When $f_i(x(t);W)=\sum_{j=1}^n w_{ij}x_j(t)$, the model recovers the Taylor model~\cite{MT:68}, in which all individuals’ opinions converge to the external input $x^*$ whenever $\V_p$ is non-empty and $\G(W)$ is strongly connected. In contrast, under a median-based formulation, i.e., with $f_i(x(t);W)=\Med_i(x(t);W)$, the group’s behavior becomes more realistic: Whether opinions converge to $x^*$ now depends on how many individuals are manipulated and where they are located in the network. The following proposition provides a necessary and sufficient graph-theoretic condition for such pinning controllability. 
\begin{proposition}[Pinning controllability]\label{thm:pinning-consensus}
The system~\eqref{eq:prejudiced-dynamic} with $f_i(x(t);W)=\Med_i(x(t);W)$ achieves asymptotic consensus at the opinion $x^*$ for any $x(0) \in R^n$ if and only if every cohesive set in $\G(W)$ contains at least one individual in $\V_p$.
\end{proposition}
The result above coincides with that of~\cite{RZ-ZL-GC-WM:24} obtained under a discrete-time formulation. However, a different proof technique is required here (see Appendix~\ref{append:proof-thm:pinning-consensus}), as our model is continuous-time. Proposition~\ref{thm:pinning-consensus} enables a numerical investigation of an important question: how does a group’s resilience to external opinion manipulation depend on its underlying influence network structure? We focus on two structural metrics: link density and clustering coefficient.

To examine the effect of link density, we generate Erdős–Rényi (ER) networks with $N=30$ nodes and vary the link probability parameter over ${0.1, 0.2, \dots, 0.6}$. To examine the effect of clustering, we generate Watts–Strogatz (WS) small-world networks with $n=30$ nodes, fix the average out-degree at 6, and vary the edge rewiring probability, which is negatively correlated with clustering coefficient, over ${0, 0.1, \dots, 1}$. For each parameter value, 150 independent simulations are performed. In each simulation, a random graph with the specified parameters is generated, edges are assigned random weights and row-normalized, and all cohesive sets are enumerated. We then identify the minimal number of manipulated nodes required such that every cohesive set contains at least one of them. 

\begin{figure}[htbp]
    \centering
    \includegraphics[width=1\linewidth]{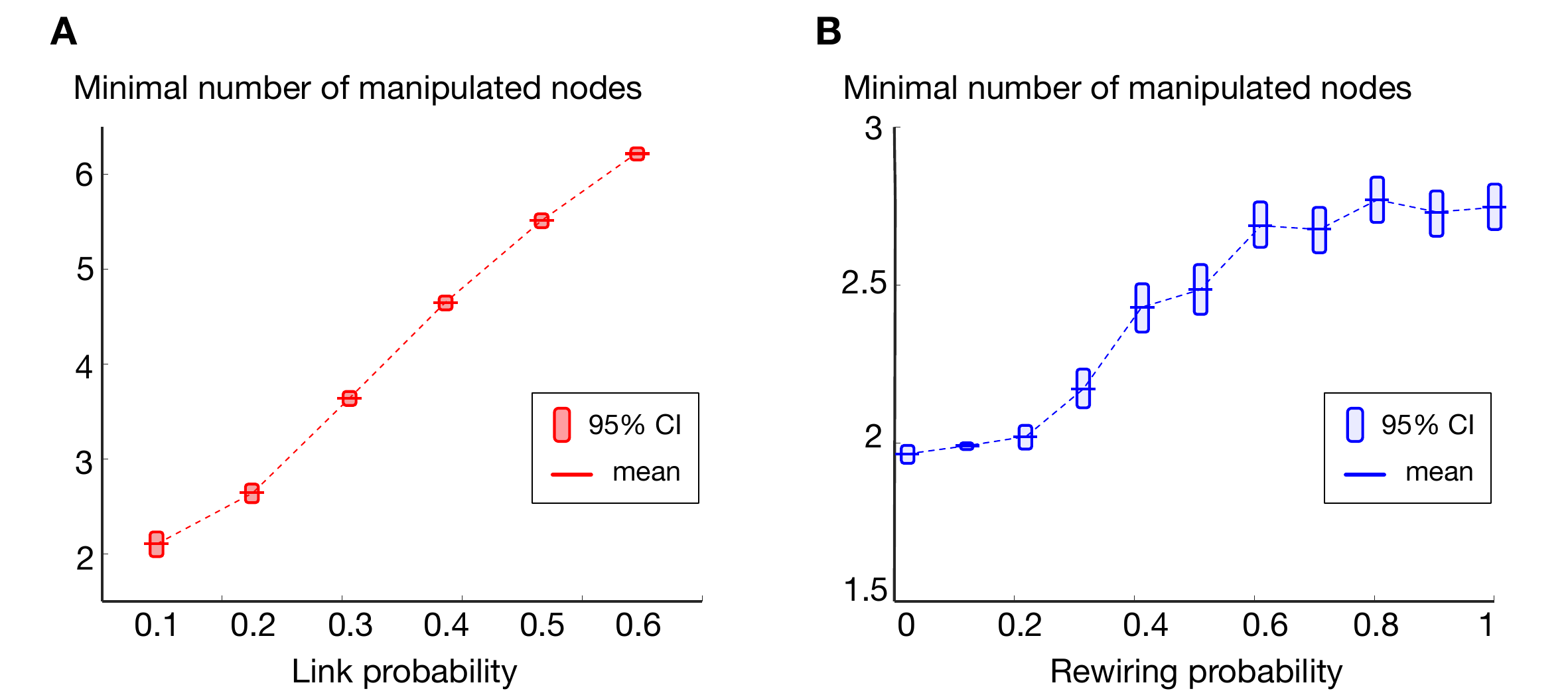}
    \caption{How network structure affects the minimal number of externally manipulated nodes required to steer the CTWM opinion dynamics to a prescribed consensus opinion. Panel~A illustrates the effect of link density, based on simulations over Erdős–Rényi (ER) random networks. Panel~B illustrates the effect of clustering coefficient (negatively correlated with rewiring probability), based on simulations over Watts–Strogatz (WS) small-world networks. In both plots, each data point represents the sample mean of 150 independent simulations under the same parameter setting, and the boxes denote 95\% confidence intervals computed from the sample standard deviation.}
    \label{fig:belt}
\end{figure}

The dependence of this minimal number on edge probability in ER networks and rewiring probability in WS networks is shown in Fig.\ref{fig:belt}A and~B, respectively. The simulation results convey a clear message that denser and less clustered social networks exhibit greater resilience to external opinion manipulation.

\section{Conclusions}
In conclusion, this paper has developed a continuous-time weighted-median opinion dynamics model that captures opinion compromise through the weighted-median mechanism. We have established key mathematical properties of the model, including the existence and uniqueness of solutions and the full characterization of equilibria. Furthermore, we have derived both necessary and sufficient graph-theoretic conditions for consensus and provided sufficient conditions for disagreement. These results offer theoretical foundations for designing strategies in social networks to promote agreement or interpret persistent divergence. Finally, by examining the role of cohesive sets in control, we reveal a structural implication of the weighted-median mechanism: sparse and clustered networks are significantly more controllable, whereas dense or well-mixed networks exhibit stronger resistance to external opinion manipulation via pinning control.

\appendix
\section{Proof of Lemma~\ref{lem:CT-from-eq-set-to-single-eq}}\label{append:proof-lem:CT-from-eq-set-to-single-eq}

Suppose $x^*,\,y^*$ are both in $\omega(x_0)\subseteq \chi$, and $x^*\neq y^*$. There exists $\epsilon>0$ such that 
\begin{align*}
\inf_{z\in B(x^*,\epsilon)}\lVert y^*-z\rVert > \epsilon,
\end{align*}
where $B(x^*,\epsilon) = \big\{ z\in D \,\big|\, \lVert z-x^*\rVert < \epsilon \big\}$. Since $x^*$ is stable, for the $\epsilon$ given above, there exists $\delta>0$ such that, for any $z$ satisfying $\lVert z - x^* \rVert< \delta$, $\phi(t,z)\in B(x^*,\epsilon)$ for any $t\ge 0$. Moreover, since $x^*$ is an $\omega$-limit point of $x_0$, there exists a time sequence $\{t_p\}_{p\in \mathbb{N}}$ such that $t_p\to \infty$ as $p\to \infty$, and $\lim_{p\to \infty}\phi(t_p,x_0) = x^*$. As a consequence, there exists $P$ such that 
$\lVert \phi(t_P,x_0) - x^* \rVert<\delta$, which implies that $\phi(t,x_0)\in B(x^*,\epsilon)$ and thereby $\lVert \phi(t,x_0)-y^* \rVert>\epsilon$ for any $t\ge t_P$. Therefore, $y^*$ cannot be an $\omega$-limit point of $x_0$. This concludes the proof.
\hfill \qed

\section{Proof of Lemma~\ref{lem:refine-omega-limit-set}}\label{append:proof-lem:refine-omega-limit-set}

Since $f(x)$ is globally Lipschitz, for any given $x\in G$, the solution $\phi(t,x)$ exists and is unique for any $t\ge 0$, and $\phi(t,x)$ is a continuous function of $x$ for any fixed $t$.
   
   Since $B$ is an open set and there exists $T>0$ satisfying $\phi(T,x)\in B$, there exists $\epsilon>0$ such that, for any $z$ satisfying $\lVert z-\phi(T,x) \rVert<\epsilon$, we have $z\in B$. Here $\lVert \cdot \rVert$ is some norm in $\mathbb{R}^n$. Moreover, since $B\cup \partial B$ is positively invariant, $\phi(t,z)\in B\cup \partial B$ for any $t\ge 0$. Additionally, since $\phi(t,x)$ is a continuous function of $x$, for the $\epsilon>0$ given above, there exists $\xi>0$ such that, for any $\hat{x}$ satisfying $\lVert \hat{x}-x\rVert<\xi$, $\lVert \phi(T,\hat{x})-\phi(T,x)\rVert<\epsilon$. Therefore, $\phi(t+T,\hat{x})\in B\cup \partial B$, for any $t\ge 0$.
   
   Suppose there exists $y\in G$ such that $x\in \omega(y)$, then there exists a sequence $\{t_k\}_{k\in \mathbb{N}}$ such that $\lim_{k\to \infty}t_k=\infty$ and $\lim_{k\to \infty} \phi(t_k,y) = x$, which implies that there exists $K\in \mathbb{N}$ such that, for any $k\ge K$, $\lVert \phi(t_k,y)-x \rVert<\xi$. Therefore, 
   \begin{align*}
   \lVert \phi\big(T,\phi(t_K,y)\big)-\phi(T,x) \rVert &= \lVert \phi(T+t_K,y)-\phi(T,x) \rVert \\&<\epsilon.
   \end{align*}  
As a consequence, $\phi(t+T+t_K,y)\in B\cup \partial B$ for any $t\ge 0$, and thereby $\lVert \phi(t,y)-x \rVert \ge \textup{dist}(x,B\cup \partial B)>0$ for any $t\ge T+t_K$, which contradicts the assumption that $x\in \omega(y)$. This concludes the proof.\hfill \qed

\section{Proof of Proposition~\ref{prop:positive_invar_cohesive}}\label{append:proof-prop:positive_invar_cohesive}
Obviously, $X_{\C,[a,b]}$ is a convex and closed set. By Theorem~\ref{thm:Med-NonExpanding}, $F(x)=\Med\big( x;W \big)$ is a Lipschitz continuous vector field. Then, given any $x\in X_{\C,[a,b]}$, let $\underline{c}_t = \min_{k\in \C} x_k$ and $\overline{c}_t = \max_{k\in \C} x_k$. For any $j\in \C$, since $\sum_{k\notin \C} w_{jk}< \frac{1}{2}$, according to Definition~\ref{def:weighted-median-in-general}, we have that 
\begin{align*}
a \leq\underline{c}_t \le \Med_j\big( x;W \big) \le \overline{c}_t\leq b.
\end{align*}
Thus $F(x)\in X_{\C,[a,b]}.$
Consequently, $X_{\C,[a,b]}$ is positively invariant by Proposition \ref{prop:fwd_inv_generic_short}.
Particularly, for any $x\in \mathbb{R}^n$, if we take $a=\min_{k\in \C} x_k$, $b=\max_{k\in \C} x_k$, we have then 
$$\phi(t,x)\in X_{\C,[a,b]},\quad \forall t\geq 0,$$
which indicates that $\max_{k\in \C}\phi_k(t,x)$ is non-increasing and $\min_{k\in \C} \phi_k(t,x)$ is non-decreasing, for any $t\ge 0$.
This concludes the proof.\hfill \qed

\section{Proof of Theorem~\ref{thm:CT-convergence}}\label{append:proof-thm:CT-convergence}

The key to the proof of Theorem~\ref{thm:CT-convergence} lies in analyzing the behaviors of states outside $\Xeq$. The sketch of the proof begins by showing that every state outside $\Xeq$ must belong to one of two specific classes. Next, we show that the system will steer away from states belonging to these specific classes. Finally, we leverage this fact to show that the $\omega$-limit set of the system is contained within $\Xeq$. Combining this result with the stability of equilibria, we establish the system’s convergence to an equilibrium in $\Xeq$.

\begin{figure}
\begin{center}
\includegraphics[width=1.0\linewidth]{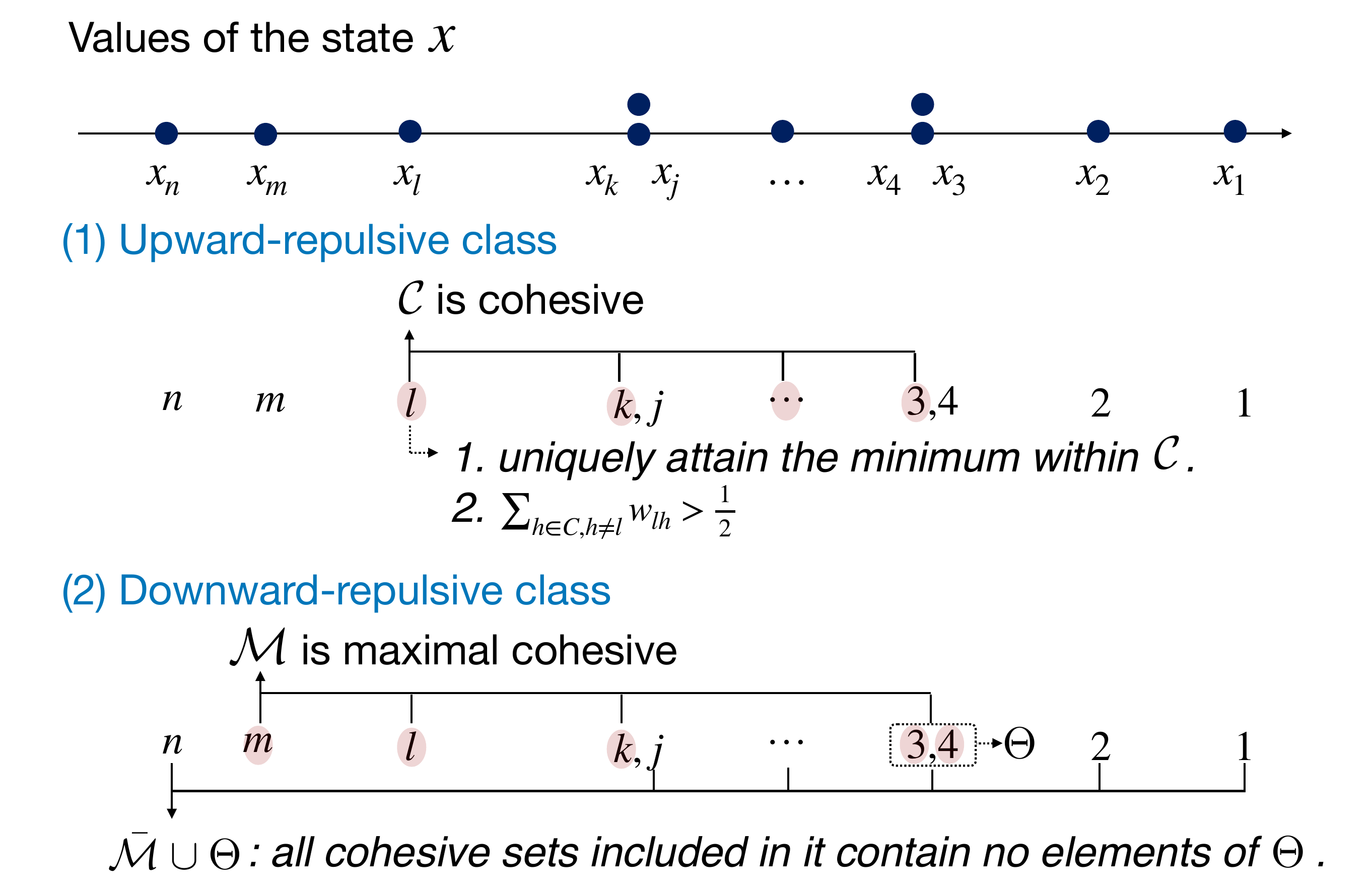}
\caption{An illustration of the two repulsive classes. For simplicity, the vector $x\in\mathbb{R}^n$ is assumed to satisfy $x_{1}\geq x_{2}\geq \cdots \geq x_{n}$, which are distributed  the axis. (1) Upward-repulsive class: suppose $\mathcal{C}=\{3,\dots,k,l\}$ is a cohesive set. The minimum within $\mathcal{C}$ is uniquely attained at $l$, and it holds that $\sum_{h\in \mathcal{C},,h\neq l} w_{lh} > \tfrac{1}{2}$. Under these conditions, $x$ belongs to the upward-repulsive class.  (2) Downward-repulsive class: suppose $\mathcal{M}=\{3,4,k,m,l\}$ is a maximal cohesive set. Let $\Theta = \arg\max_{i\in \mathcal{M}} x_{i}=\{3,4\}$. No cohesive set contained in $\Theta \cup \bar{\mathcal{M}}=\{1,2,3,4,\cdots,j,n\}$ includes any element of $\Theta$. Under these conditions, $x$ belongs to the downward-repulsive class.
}\label{fig:two-states}
\end{center}
\end{figure}

To formalize the above proof sketch, we now introduce several lemmas. The following lemma classifies states outside $\Xeq$ into two classes. As both classes will subsequently be shown to exhibit repulsive behaviour under the system dynamics in Lemma~\ref{lem:min_cohesive} and Lemma~\ref{lem:max_cohesive_max_noco}, we refer to them as upward-repulsive class and downward-repulsive class for convenience. An illustration of the two classes is presented in Fig~\ref{fig:two-states} based on Lemma~\ref{lem:noXeq->class_1_2}.

\begin{lemma}\label{lem:noXeq->class_1_2}
Let $\Xeq$ be the set of all equilibria, as defined in Theorem~\ref{thm:CT-EqSet} .  If $x\not\in \Xeq$, then $x$ is in at least one of the following two classes of states:
    \begin{itemize}
    \item Upward-repulsive class: A state $x$ belongs to this class if there exists a cohesive set $\C$ such that the minimum component within $\C$ is uniquely attained at some index $i^* \in \mathcal{C}$, i.e., $\{i^*\} = \argmin_{j \in \mathcal{C}} x_j$, and this $i^*$ satisfies $\sum_{j\in \C,j\neq i^*} w_{i^*j} >\frac{1}{2}$;
    \item Downward-repulsive class: A state $x$ belongs to this class if there exists a maximal cohesive set $\M$ such that no cohesive set included in $\Theta \cup \bar \M$ contains elements of $\Theta$, where $\Theta = \arg\max_{i\in \M} x_{i}$.
    \end{itemize}
\end{lemma}

\emph{Proof:} 
If $x\not\in \Xeq$, then there exists $r\leq \Ndiff-1$ such that $\order{\leq k} (x)$ is maximally cohesive for every $k<r$ but $\order{\leq r} (x)$ is not. This situation includes the trivial case where $r=1$.

    We let then $S\subseteq \order{\leq r} (x)$ be the union of all cohesive sets included in $\order{\leq r} (x)$, including $\order{<r}(x)$ since it is by assumption maximally cohesive, but also possibly nodes of $\order{r}(x)$. Since the union of cohesive sets is cohesive (see Lemma~\ref{lem:properties-cohesive-expansion}), $S$ is cohesive. We consider two situations.

    1) $S$ is not maximally cohesive. In that case there is a node $i^*\not \in S$ such that $\sum_{j\in S} w_{i^*j}>\frac{1}{2}$, which implies that $S\cup \{i^*\}$ is cohesive. This in turns implies that $i^*\not \in \order{\leq r} (x)$ for otherwise it would have been included in $S$, and therefore that $x_{i^*}< x_i, \forall i\in S$. We are thus in the situation of upward-repulsive class with $C=S\cup \{i^*\}$.
%\noteus{that case can only happen if there are some nodes from $\order{r}(x)$ in $S$, otherwise the assumptions are contradicted.}

 %\textcolor{red}{HY: A little confusing:by assumption, $\order{<r}(x)$ is unnecessarily to be maximally cohesive }

    2) Otherwise, $S$ is maximally cohesive, and so is thus $\bar S$ by Lemma \ref{lem:properties-cohesive-expansion}. Moreover, $\bar S$ does have a non-empty intersection with $\order{\leq r} (x)$, for otherwise we would have a maximally cohesive $\order{\leq r} (x)=S$. Hence $\Theta := \arg\max_{i\in \bar S} x_i$ is a subset of $\order {\leq r}(x)$. Moreover, by construction, there is no cohesive set included in $S\cup \Theta$ that contains elements of $\Theta$, for otherwise these elements would have been included in $S$. Hence we are in downward-repulsive class with $\M = \bar S$.
\hfill \qed

Next, we will show that trajectories starting from these two classes of states eventually enter positively invariant sets that are at non-zero distances from their initial points. For convenience, we state first the following rather obvious lemma, which directly follows from the Lipschitz continuity of $\Med(x,W)-x$. 

\begin{lemma}\label{lem:obvious_growth}
    Suppose $\Med_i(x(t),W) > x_i(t)$. Then there exists a time $t'>t$ at which $x_i(t') > x_i(t)$, and $x_i(t')< x_j(t')$ for every $j$ such that $x_j(t)>x_i(t)$, i.e. $x_i$ will increase but did not exceed any agent that had a larger opinion.    
\end{lemma}

For the upward-repulsive class, Lemma~\ref{lem:min_cohesive} implies that the minimum opinion within a cohesive set is lifted over time, while for the downward-repulsive class, lemma~\ref{lem:max_cohesive_max_noco} indicates the maximum opinion within a maximal cohesive set is lowered. 

\begin{lemma}[Upward behaviour]\label{lem:min_cohesive}
%    Let  $\C$ be a cohesive set. Suppose that vector $x_0\in \mathbb{R}^n$ has the following properties:
%    \begin{enumerate}
%    \item $\min_{j\in \C}x_{0,j}$ is reached by one single node $i^*\in \C$;
%    \item $\sum_{j\in \C,j\neq i^*} w_{i^*j} >\frac{1}{2}$.
%\end{enumerate}  
%    Then there is an $\epsilon>0$ and a time $\tau$ after which  $$\phi_j(t,x_0)\geq x_{i^*}+\epsilon, \forall t\geq \tau, j\in \C$$

%\comjh{alternative}
Suppose that vector $x_0\in \mathbb{R}^n$ is in upward-repulsive class as defined in Lemma~\ref{lem:noXeq->class_1_2}. Then there exists $\epsilon>0$ and a time $\tau$ such that  
  $$  
\phi(\tau,x_0)\in X^a_{x_0}:=\{x: x_i\geq \min_{j\in \C}x_{0,j} +\epsilon , \forall i\in \C\},
$$
where $X^a_{x_0}$ is a positively invariant set and is at a positive distance from $x_0$.

%\noteus{need a better name for the set, especially since it does not depend only on $x_0$. Or we could say, there exists a positive invariant set $X^a(x^0)$ at positive distance from $x_0$ s.t.}
%\textcolor{red}{HY: I prefer the former way. And I add a remark after the proof. The later expression is a little indirect and would require us to restate its equivalent form in the proof.    }  
\end{lemma}

\emph{Proof:} 
Since $\sum _{j\in \C,j\neq i^*} w_{i^*j}>\frac{1}{2}$ and $x_{0,j}>x_{0,i^*}$ for all $j\in \C$, we have $\Med_{i^*}(x_0,W)>x_{i^*}$. Then it follows from Lemma \ref{lem:obvious_growth} that there is a time $\tau$ at which  $\phi_{i^*}(x_0,\tau) > x_{i^*}$ while keeping $\phi_{i^*}(x_0,\tau)\leq \phi_{j}(x_0,\tau)$ for every other $j\in \C$. Thus we can take $\epsilon=\phi_{i^*}(x_0,\tau) - x_{i^*}>0$ to construct $X^a_{x_0}$. The rest of the result follows then directly from Proposition \ref{prop:fwd_inv_generic_short} on positive invariance for cohesive sets. \hfill \qed

Note that we could build a parallel version of this result if $i^*$ takes the maximal value, switching the sign of $\epsilon$ and of the inequalities. Additionally, it should be noted that $X^a_{x_0}$ is not a function of $x_0$, but rather a notation used to represent a positively invariant set associated with $x_0$ when it is in upward-repulsive class. Similarly, the usage of $X^b_{x_0}$ in Lemma~\ref{lem:max_cohesive_max_noco} follows the same logic.  

%\noteus{The next proposition is probably the most obstruse, maybe a graph would help. Alternatively, if we first show Proposition \ref{lem:noXeq->class_1_2}, its context my be clearer.\\ Below I'm using the notion $\bar \M$ to denote the set $\V\setminus \M$. If this Proposition is used, the notation should be made uniform in one way or another. I'm also proposing two versions for the proof.}
%\textcolor{red}{HY: 1.	I have added Fig~\ref{fig:noXeq-two-class} to demonstrate the two states when \(x \notin \Xeq\). This figure is designed for Proposition~\ref{lem:noXeq->class_1_2}. Do we need another graph to illustrate Proposition~\ref{lem:max_cohesive_max_noco}?\\2.	The meaning of $\bar{M}$ is clarified in the table of notations.\\	3.	The first version is adopted in the current manuscript.\\}

\begin{lemma}[Downward behaviour]\label{lem:max_cohesive_max_noco}
    Suppose that vector $x_0\in \mathbb{R}^n$ is in downward-repulsive class as defined in Lemma~\ref{lem:noXeq->class_1_2}.    
%    Then there exists $\epsilon >0$ and a time $\tau$ after which 
%    $$\phi_j(x,t) \leq \max_{i\in \M} x_i  - \epsilon, \forall j\in \M, t\geq \tau.$$
%    \comjh{alternative check indices}
Then there exists $\epsilon >0$ and a time $\tau$ at which $\phi(t,x_0)$ enters the positively invariant set
$$
X^b_{x_0}:=\{x: x_j\leq \max_{i\in \M} x_{0,i}  - \epsilon, \forall j\in \M\}
$$
which is at a positive distance from $x_0$.
\end{lemma}    

\emph{Proof:} First, $\bar \M$ is a maximal cohesive set by Lemma \ref{lem:properties-cohesive-expansion}. Observe then there has to exists a node $i^*\in \Theta\subseteq \M$ for which $\sum_{j\in \M\setminus \Theta} w_{i^*j} > \frac{1}{2}$. Indeed, otherwise we would have $\sum_{j\in \bar M \cup  \Theta} w_{ij} \geq \frac{1}{2}$ for every $i\in \Theta$, and $\Theta \cup \bar \M$ would be a cohesive set, in contradiction with our assumption.
Since all $x_{0,j} $ for $j\in \M\setminus \Theta$ are by definition of $\Theta$ smaller than $x_{0,i^*}$, we have $\Med_{i^*}(x_0,W)<x_{0,i^*}$. From Lemma \ref{lem:obvious_growth} (in reverse), it follows again that there is a time $\tau$ at which  $\phi_{i^*}(x_0,\tau) < x_{0,i^*}$ while keeping $\phi_{j}(x_0,\tau)< \phi_{i^*}(x_0,\tau)$ for every other $j\in \M\setminus \Theta$. Let $\Theta'$ be the set of nodes $i\in \M$ such that  $\phi_{i}(x_0,\tau)= x_{0,i^*}$. Clearly $\Theta'$ is a proper subset of $\Theta$, since it does not contain $i^*$, and cannot contain any element of $\M\setminus \Theta$ either, as we have seen these nodes have opinions   $\phi_i(x_0,tau)=x_{0,i^*}$. Moreover, we still have $\phi_j(x_0,\tau)\leq x_{0,i^*}$ for every $j\in \M$ by Proposition \ref{prop:fwd_inv_generic_short}.

We now proceed by recurrence on the number of elements in $\Theta$. If $|\Theta|=1$, then $\Theta'$ is empty, i.e. $\phi_{i}(x_0,\tau) < x_{0,i^*}$ for every $i\in \Theta$, and the result holds. The same holds if $|\Theta|>1$ but $\Theta'$ is empty. If $|\Theta|>1$ and $\Theta'$ is not empty, observe that the assumptions of the proposition hold at $\phi(x_0,\tau)$ for this new smaller $\Theta'$ and for the same maximum value $\max_{i\in \M} x_{0,i}\phi_i(x_0,\tau)=x_{i^*}$, and the result holds by iterating the above augment. 

Finally, the positive invariance of the set follows from Proposition \ref{prop:fwd_inv_generic_short}.
\hfill\qed

With the above preparations, we are ready to present a complete proof of Theorem~\ref{thm:CT-convergence}.

\emph{Proof of Theorem~\ref{thm:CT-convergence}:} 
First of all, Proposition \ref{prop:fwd_inv_generic_short} implies that the trajectory remains in the compact set $G=\{x:x_i\in [\min_j x_{0,j}, \max_j x_{0,j}],\forall i\in\V \}$ and therefore admits accumulation points, or $\omega$-limit points. 

Consider a state $x'\in G\setminus \Xeq$. We aim to show that $x'$ is excluded from the $\omega$-limit set of $x_0$. From Lemma \ref{lem:noXeq->class_1_2}, it follows that $x'$ must be in one of the two repulsive states and have the property described in Lemma \ref{lem:min_cohesive} or \ref{lem:max_cohesive_max_noco}. Suppose that $x'$ is in upward-repulsive class. By Lemma \ref{lem:min_cohesive}, there exists $\epsilon>0$ such that the trajectory enters a positively invariant set $X^a_{x'}$ at time $\tau$, and of course enters $G\cap X^a_{x'}$. To ensure the trajectory enters the interior of a set, as required by Lemma~\ref{lem:CT-from-eq-set-to-single-eq}, we introduce a "larger" positively invariant set
$$
\tilde X^a_{x'}:=\{x: x_i\geq \min_{j\in \C}x_{1,j} +\epsilon/2 , \forall i\in \C\},
$$
where $\C$ is the cohesive set associated with $X^a_{x'}$. Moreover, define $$G_\delta=\{x:x_i\in [\min_{j\in \V} x_{0,j}-\delta, \max_{j\in\V} x_{0,j}+\delta],\forall i\in \V \},$$ where $\delta=\max_j x_{0,j}-\min_j x_{0,j}$. For their intersection, $G_\delta\cap \tilde X^a_{x'}$ is positively invariant and \(x' \not\in G_\delta \cap \tilde{X}^a_{x'}\) . Since $G\cap X^a_{x'}$ is contained in the interior of $G_\delta\cap \tilde X^a_{x'}$, if we take $B\cap \partial B = G_\delta\cap \tilde X^a_{x'}$ and take $B$ as the interior of $G_\delta\cap \tilde X^a_{x'}$, it holds that $\phi(\tau,x')\in B$.
Therefore, using Lemma~\ref{lem:refine-omega-limit-set}, for $x'\in G_\delta$, we have $x'\not\in\omega(y)$ for any $y\in G_\delta$, particularly for $x_0\in G\subset G_\delta$.
A similar argument holds when $x'$ is in downward-repulsive class, where we take $B \cap \partial B = G_\delta \cap \tilde{X}^b(x')$.
Therefore, since the elements of $\Xeq$ are all stable equilibriums by Theorem \ref{thm:CT-EqSet}, convergence to an element of $\Xeq$ follows from Lemma~\ref{lem:CT-from-eq-set-to-single-eq}. \hfill \qed
\section{Proof of Proposition~\ref{prop:continuity-measurability}}\label{append:proof-prop:continuity-measurability}

We begin by proving the (Lipschitz) continuity of $\phi(t,x)$ with respect to $x$, for every $t$. Take two initial conditions $x,y$. The corresponding solutions satisfy 
\begin{align*}
&\frac{d}{dt}\norm{\phi(t,x)-\phi(t,y)}_2^2 
\\&=2 (\phi(t,x)-\phi(t,y))^{\top}{f(\phi(t,x))-f(\phi(t,y))}.
\end{align*}

Assuming without loss of generality (in finite dimension) that $f$ is Lipschitz continuous w.r.t. the 2 norm, we have then
\begin{align*}
&\frac{d}{dt}\norm{\phi(t,x)-\phi(t,y)}_2^2 \\
&\leq 2\norm{\phi(t,x)-\phi(t,y)}_2\norm{f(\phi(t,x))-f(\phi(t,y))}_2\\
&\leq 2L \norm{\phi(t,x)-\phi(t,y)}^2_2
\end{align*}
Hence by Gronwall's inequality, 
$$
\norm{\phi(t,x)-\phi(t,y)}_2^2 \leq e^{2Lt}\norm{x-y}_2^2,
$$
or equivalently,
$$
\norm{\phi(t,x)-\phi(t,y)}_2 \leq e^{Lt}\norm{x-y}_2.
$$

This result does not directly extend to infinite time though. Hence we now combine it with convergence and stability to characterize $\bar \phi$. Let $x_0 \in \mathbb{R}^n$ and $\epsilon > 0$ be arbitrary. We aim to show that there exists $\delta>0$ such that the trajectories starting from $B(x_0,\delta)$ all converge to an equilibrium in $B(\bar \phi(x_0), \epsilon)$. Since $f$ is Lipschits continuous, $\phi(x_0)$ must be an equilibrium and hence is stable. We take the $\delta'$ provided by the definition of stability for $\epsilon/2$ we then take a time $T$ after which $\phi(t,x_0)$ remains in $B(\bar \phi(x_0),\delta'/2)$ for all time. The existence of this $T$ follows from the convergence of $\phi(t,x)$ to $\bar \phi(x)$. Finally, we take $\delta = \frac{1}{2}\delta' e^{-LT}$. 

Now for any $y\in B(x_0,\delta)$, we have by the first part of this result
\begin{align*}
\norm{\phi(T,y)-\phi(T,x_0)}_2&\leq e^{LT}\norm{y-x_0}_2\\
&\leq e^{LT}\frac{1}{2}\delta' e^{-LT}= \frac{1}{2}\delta'
\end{align*}
%\textcolor{red}{JH:for this appendix we can most likely summarizeed the end of the proof.}
Therefore, there holds
\begin{align*}
&\norm{\phi(T,y)-\bar \phi(x_0)}\\
&\leq \norm{\phi(T,y)-\phi(T,x_0)} +      
\norm{\phi(T,x_0)-\bar \phi(x_0)}\\
& \leq \frac{\delta'}{2}+\frac{\delta'}{2} = \delta'.
\end{align*}
and hence the stability of $\bar\phi(x_0)$ implies that $\phi(T,y)$ remains at a distance at most $\epsilon$ from $\bar \phi (x_0)$, and in particular $\bar\phi(y)\in B(\bar \phi(x),\epsilon)$. This establishes the continuity of $\bar \phi$. Its measurability follows then directly since every continous function is measurable. 

\section{Proof of Proposition~\ref{thm:pinning-consensus}}\label{append:proof-thm:pinning-consensus}
To prove it, we introduce the following lemmas that characterize the monotonicity and
compression properties of the partially prejudiced system (\ref{eq:prejudiced-dynamic}).
\begin{lemma} \label{lem:prejudiced-mono}
Consider the opinion dynamics (\ref{eq:prejudiced-dynamic}). For any $a \leq u \leq b$, the set $X_{\V,[a,b]}$ (follow the definition in Proposition~\ref{prop:positive_invar_cohesive}) is positively invariant.
In particular,
\begin{enumerate}[label=\roman*.]
\item 
$\max _{i \in \mathcal{V}} x_i(t)$ is monotonically non-increasing as long as $u \leq \max_{i\in \V} x_i(t)$. (respectively, $\min _{i \in \mathcal{V}} x_i(t)$ is monotonically non-decreasing as long as $\min_{i\in \V} x_i(t)\leq u$. )

\item If there exists $T \geq 0$ such that $u \geq \max _{i \in \mathcal{V}} x_i(T)$, then $u \geq$ $\max _{i \in \mathcal{V}} x_i(t)$ for all $t \geq T$.(respectively, if there exists $T \geq 0$ such that $u \leq \min _{i \in \mathcal{V}} x_i(T)$, then $u \leq$ $\min _{i \in \mathcal{V}} x_i(t)$ for all $t \geq T$.)
\end{enumerate}

\end{lemma}

\emph{Proof:}
Since $a \leq u \leq b$, for any $x \in X_{\V,[a,b]}$, the following holds:
$$
a \mathbf{1} \leq \Gamma u + (I - \Gamma)\operatorname{Med}(x; W) \leq b \mathbf{1},
$$
which implies that
$\Gamma u + (I - \Gamma)\operatorname{Med}(x; W) \in X_{\V,[a,b]}.
$
Thus, $X_{\V,[a,b]}$ is a positively invariant set for the dynamics (\ref{eq:prejudiced-dynamic}) by Proposition~\ref{prop:fwd_inv_generic_short}.

For case (i): For any $\tau$, since $x(\tau) \in X_{\V,[a,b]}$ with $a = \min_{i \in \mathcal{V}} x_i(\tau)$, $ b = \max_{i \in \mathcal{V}} x_i(\tau)$, it follows that for all $t \geq \tau$,
$$
\min_{i \in \mathcal{V}} x_i(t) \geq a, \quad \max_{i \in \mathcal{V}} x_i(t) \leq b.
$$
As this relationship holds for any $t \geq \tau$, we can iteratively refine $[a,b]$ to construct smaller positively invariant sets. Consequently, we conclude that $\max_{i \in \mathcal{V}} x_i(t)$ is monotonically non-increasing, and $\min_{i \in \mathcal{V}} x_i(t)$ is monotonically non-decreasing.

For case (ii): Consider the set $X_{\V,[a,b]}$ with $a=\min_{i \in \mathcal{V}} x_i(T)$, $b=u$. Since this set is positively invariant, the result follows. Similarly, for the opposite bound, we consider the set \( X_{\V,[a,b]} \) with  $a = u$  and  $b = \max_{i \in \mathcal{V}} x_i(T)$ and obtain the result.
\hfill \qed

\begin{lemma}\label{lem:prejudiced-steer-all}
Consider the opinion dynamics (\ref{eq:prejudiced-dynamic}). Suppose that $\G(W)$ does not contain a cohesive set consisting of unprejudiced agents only. Then,

\begin{enumerate}[label=\roman*.]

\item %Suppose there exists $T \geq 0$ such that $\max _{i \in \mathcal{V}} x_i(t)$ is monotonically non-increasing for $t \geq T$ .
If there exist $M$ and $T_1$ such that $\max_{i\in \V_p}x_i(t)\leq M$ for all $t\geq T_1$,
then for any $\epsilon >0$, there exists $T_1^*$ such that 
$$\max_{i\in \V_u}x_i(t)\leq M+\epsilon ,\quad \forall t\geq T_1^*.$$

\item If there exists $T \geq 0$ such that  $\max_{i \in \mathcal{V}} x_i(t)$ is monotonically non-increasing for $t \geq T$ and $u\leq\max _{i \in \mathcal{V}} x_i(T)$, then for any $\epsilon>0$, there exist $T'$ such that
\begin{equation}\label{eq:compress-max-prejudiced}
    x_i(t)\leq u+\epsilon, \quad i \in \mathcal{V}, t\geq T'.
\end{equation}
\end{enumerate}
\end{lemma}
% Suppose there exists $T \geq 0$ such that $\min _{i \in \mathcal{V}} x_i(t)$ is monotonically non-decreasing for $t \geq T$ .  If there exist $M$ and $T_1$ such that $\min_{i\in \V_p}x_i(t)\geq M$ for all $t\geq T_1$,then for any $\epsilon >0$, there exists $T_1^*$ such that $$\min_{i\in \V_u}x_i(t)\geq M-\epsilon ,\quad \forall t\geq T_1^*.$$
\emph{Proof:}
(i) Let $\epsilon>0$.
By the condition of this lemma, neither $\mathcal{V}_u$ nor any of its subsets is not a cohesive set. From the definition of cohesive sets, there exists an agent $k_1 \in \mathcal{V}_u$ such that
$
\sum_{j \in \mathcal{V}_u} w_{k_1 j} < \frac{1}{2}.
$
This implies
$$
\sum_{j \in \mathcal{V}_p} w_{k_1 j} = 1 - \sum_{j \in \mathcal{V}_u} w_{k_1 j} > \frac{1}{2}.
$$
Similarly, as $\mathcal{V}_u \setminus \{k_1\}$ is not a cohesive set, there exists another agent $k_2 \in \mathcal{V}_u \setminus \{k_1\}$ such that
$
\sum_{j \in \mathcal{V}_u \setminus \{k_1\}} w_{k_2 j} < \frac{1}{2},
$
which leads to
$$
w_{k_2 k_1} + \sum_{j \in \mathcal{V}_p} w_{k_2 j} = 1 - \sum_{j \in \mathcal{V}_u \setminus \{k_1\}} w_{k_2 j} > \frac{1}{2}.
$$
By repeating this process, we can sequentially identify agents $k_3, k_4, \dots, k_{n_2}$ such that for $i \in \{3, \dots, n_2\}$, $k_i \in \mathcal{V}_u \setminus \{k_1, \dots, k_{i-1}\}$ and
$$
\sum_{1 \leq l \leq i-1} w_{k_i k_l} + \sum_{j \in \mathcal{V}_p} w_{k_i j} > \frac{1}{2}.
$$

Next, we use induction to prove the claim that for any $i\in \{1,\cdots,n_2\}$, there exists $\tau_i$ such that
\begin{align*}
x_{k_j}(t)\leq M+\frac{j}{n_2}\epsilon,\forall t\geq \tau_i, \forall j\in \{1,\cdots,i\}.
\end{align*} 
When $i=1$, the assumption $\max_{i\in \V_p}x_i(t)\leq M$ implies that for all $ t\geq T_1$
\begin{align*}
\dot x_{k_1}(t)&= \operatorname{Med}_{k_1}(x(t) ; W)-x_{k_1}(t)\\
&\leq \max_{j\in \V_p} x_j(t)- x_{k_1}(t)\\
&\leq M-x_{k_1}(t).
\end{align*}
By the comparison principle, we obtain $x_{k_1}(t)\leq M+e^{-(t-T_1)}(x_{k_1}(T_1)-M),\forall t\geq T_1.$
Obviously there exists $\tau_1\geq T_1$ such that 
$x_{k_1}(t)\leq M+\frac{1}{n_2}\epsilon,\forall t\geq \tau_1.$
Then, assume that the inequality holds for $i=i^*-1$. 
That is, there exist $\tau_{i^*-1}$ such that
\begin{align*}
x_{k_j}(t)\leq M+\frac{j}{n_2}\epsilon,\forall t\geq \tau_{i^*-1}, \forall j\in \{1,\cdots,i^*-1\}.
\end{align*} 
From this assumption, it can be directly obtained that for all $t\geq \tau_{i^*-1}$,
\begin{align*}
\dot x_{k_i{^*}}(t)&= \operatorname{Med}_{k_{i^*}}(x(t) ; W)-x_{k_{i^*}}(t)\\
&\leq \max _{j \in \mathcal{V}_p \cup \{k_1,\cdots,k_{i^*-1}\}} x_j(t) -x_{k_{i^*}}(t)\\
&\leq M+\frac{i^*-1}{n_2}\epsilon -x_{k_{i^*}}(t).
\end{align*}
which follows that \begin{align*}
 x_{k_i{^*}}(t)\leq &M+\frac{i^*-1}{n_2}\epsilon \\&+e^{-(t-\tau_{i^*-1})}(x_{k_1}(\tau_{i^*-1})-M-\frac{i^*-1}{n_2}\epsilon).   
\end{align*}
Then, there exists $\tau_{i^*}\geq \tau_{i^*-1}$ such that
\begin{align*}
x_{k_{i^*} }(t)\leq M+\frac{i^*}{n_2}\epsilon,\forall t\geq \tau_{i^*}.
\end{align*} Up to now, we have proved the case when $i=i^*$. By induction, the inequality holds for all $i \in \{1, \dots, n_2\}$. Taking $T_1^* = \tau_{n_2}$, we conclude that:
$$
x_{k_j}(t) \leq M +\epsilon, \quad \forall t \geq T_1^*, \, \forall j \in \{1, \dots, n_2\}.
$$
This completes the proof of (i).

(ii) Let $\gamma_{\min}=\min_{i\in\V_p}\gamma_i$ and define $M_0:=\max_{j\in\V}x_j(T)$. For any $i\in \V_p$ and all $t\ge T$, we have
\begin{align*}
\dot{x}_i(t) & = \gamma_i u+\left(1-\gamma_i\right) \operatorname{Med}_i(x(t) ; W)-x_i(t) \\
& \leq \gamma_i u+\left(1-\gamma_i\right) M_0 -x_i(t)\\
& \leq \gamma_{\min} u +\left(1-\gamma_{\min}\right) M_0 -x_i(t),
\end{align*}
where the first inequality follows from $\operatorname{Med}_i(x(t) ; W)\leq \max_{j\in\V}x_j(t)\le M_0$ for $t\ge T$.
By the comparison principle, it can be obtained that for $ i  \in \mathcal{V}_p$ and $ t \geq T$:
\begin{align*}
x_i(t)\le &\bigl[\gamma_{\min}u+(1-\gamma_{\min})M_0\bigr]
\\&+\bigl(x_i(T)-\gamma_{\min}u-(1-\gamma_{\min})M_0\bigr)e^{-(t-T)}.
\end{align*}
Thus, there exists $T_1\geq T$ such that
\begin{align*}
x_i(t)\leq \gamma_{\min}u+(1-\gamma_{\min})M_0 +\frac{1}{2}\epsilon,\forall i  \in \mathcal{V}_p, t \geq T_1.
\end{align*}
By Lemma~\ref{lem:prejudiced-steer-all} (i), this bound on $\max_{i\in \V_p}x_i(t)$ implies the existence of $T^*_1$ such that
\begin{align*}
x_i(t)\leq \gamma_{\min}u+(1-\gamma_{\min})M_0 +\epsilon,\forall i  \in \mathcal{V}, t \geq T_1^*.
\end{align*}

Now choose $\epsilon:=\frac{\gamma_{\min}}{2}\,(M_0-u)\ge 0
$ to obtain the bound
\[
\max_{i\in\V}x_i(t)\le \Bigl(1-\tfrac{\gamma_{\min}}{2}\Bigr)M_0+\tfrac{\gamma_{\min}}{2}\,u
=: M_1,\quad t\ge T_1^*.
\]
We now restart the same argument at time $T_1^*$ (the assumptions remain valid) with
$M_0$ replaced by $M_1$, and obtain $T_2^*\ge T_1^*$ such that
\[
\max_{i\in\V}x_i(t)\le M_2:=\Bigl(1-\tfrac{\gamma_{\min}}{2}\Bigr)M_1+\tfrac{\gamma_{\min}}{2}\,u,
\quad t\ge T_2^*.
\]
Iterating yields the recursion
\[
M_{k+1}=\Bigl(1-\tfrac{\gamma_{\min}}{2}\Bigr)M_k+\tfrac{\gamma_{\min}}{2}\,u,
\qquad k=0,1,2,\dots,
\]
and thus the closed form
\[
M_k=u+(M_0-u)\Bigl(1-\tfrac{\gamma_{\min}}{2}\Bigr)^k.
\]
Therefore, there exists an increasing sequence of times $\{T_k^*\}$ with
\[
\max_{i\in\V}x_i(t)\le u+(M_0-u)\Bigl(1-\tfrac{\gamma_{\min}}{2}\Bigr)^k,
\quad t\ge T_k^*.
\]
Since $1-\tfrac{\gamma_{\min}}{2}\in(0,1)$, the right-hand side converges geometrically to $u$, establishing the desired inequality in~\eqref{eq:compress-max-prejudiced}. \hfill \qed

A similar result holds for the minimum values of  $x_i(t)$ , but for brevity, we do not explicitly state the symmetric case.
With all the preparations, now we begin to prove Proposition~\ref{thm:pinning-consensus}.

\emph{Proof of Proposition~\ref{thm:pinning-consensus}:}

(Sufficiency): We consider two cases for the system behaviour:

Case I: The state of system (\ref{eq:prejudiced-dynamic}) satisfies $\min _{i \in \mathcal{V}} x_i(t)<$ $u<\max _{i \in \mathcal{V}} x_i(t)$ for any $t \geq 0$. From Lemma~\ref{lem:prejudiced-mono}(i), we get $\max _{i \in \mathcal{V}} x_i(t)$ and $\min _{i \in \mathcal{V}} x_i(t)$ are monotonically non-increasing and non-decreasing respectively for $t \geq 0$. Because $\gamma_{\min } \in(0,1]$, by Lemma~\ref{lem:prejudiced-steer-all} (ii) we have
\begin{equation}\label{eq:proof:thm:consensus}
\lim _{t \rightarrow \infty} x_i(t)=u, i \in \mathcal{V}
\end{equation}
Case II: There exists $T \geq 0$ such that $u \leq \min _{i \in \mathcal{V}} x_i(T)$ or $u \geq \max _{i \in \mathcal{V}} x_i(T)$. Since the analysis of these two cases are similar, we assume $u \leq \min _{i \in \mathcal{V}} x_i(T)$ without loss of generality. By Lemma~\ref{lem:prejudiced-mono}, we have
$$
x_i(t)-u \geq 0, \quad \forall i \in \mathcal{V}, t \geq T
$$
and $\max _{i \in \mathcal{V}} x_i(t)$ is monotonically non-increasing for $t \geq T$. Combining this with Lemma~\ref{lem:prejudiced-steer-all}(ii), for all agents we have consensus as (\ref{eq:proof:thm:consensus}).

By putting together Case I and II, for any $\boldsymbol{x}(0) \in$ $\mathbb{R}^n$, the system (\ref{eq:prejudiced-dynamic}) achieves asymptotic consensus at $u\vectorones[n]$.

(Necessity): We prove necessity by contradiction. Assume there exists a cohesive set $\mathcal{M} \subset \mathcal{V}$ consisting of unprejudiced agents only. For all $i \in \mathcal{M}$ and $t \geq 0$, it follows from the definition of cohesive sets that:
$$
\min _{j \in \mathcal{M}} x_j(0) \leq x_i(t) \leq \max _{i \in \mathcal{M}} x_j(0).
$$
Let $a$ be a real number satisfying $a>u$. Choose $x_i(0)=a$ for any $i \in \mathcal{V}$. Since $\mathcal{M} \subset \mathcal{V}_u$, we have
$$
x_i(t)=a, \quad \forall i \in \mathcal{M}, t \geq 0.
$$
Then by (\ref{eq:prejudiced-dynamic}), for any $i \in \mathcal{V}_p$ we have for $t\geq 0$
$$
\begin{aligned}
\dot x_i(t) & =\gamma_i u+\left(1-\gamma_i\right) \operatorname{Med}_i(x(t) ; W)-x_i(t) \\
& \leq \gamma_i u+\left(1-\gamma_i\right) a-x_i(t),
\end{aligned}
$$
where the inequality follows from Lemma~\ref{lem:prejudiced-mono} (ii). 
Thus, we have $x_{i}(t)\leq \gamma_i u+\left(1-\gamma_i\right) a +e^{-t}(\gamma_i a-\gamma_i u),\forall t\geq 0,\forall i\in \V_p.$
By this we have obtained a contradiction and thus the consensus can not be reached.
\hfill\qed

\bibliographystyle{plain}        % Include this if you use bibtex 
\bibliography{alias,WM,Main,New,HY_Add,ref_jh_median} 

@article{VA-FB-AKS:17,
  title={Polar Opinion Dynamics in Social Networks},
  author={Victor Amelkin and Francesco Bullo and Ambuj K. Singh},
  journal={IEEE Transactions on Automatic Control},
  year={2017},
  volume={62},
  pages={5650-5665},
  url={https://api.semanticscholar.org/CorpusID:21249511}
}

@article{FG-EO-SB:19,
  title={The dynamics of opinion expression},
  author={Felix Gaisbauer and Eckehard Olbrich and Sven Banisch},
  journal={Physical review. E},
  year={2019},
  volume={102 4-1},
  pages={
          042303
        },
  url={https://api.semanticscholar.org/CorpusID:209515661}
}

@article{MC-ACL-BP:15,
  title={A nonlinear model of opinion formation on the sphere},
  author={Marco Caponigro and Anna Chiara Lai and Benedetto Piccoli},
  journal={Discrete and Continuous Dynamical Systems},
  year={2015},
  volume={35},
  pages={4241-4266},
  url={https://api.semanticscholar.org/CorpusID:13849694}
}

@article{QZ-GK-HZ:20,
  title={Opinion dynamics in finance and business: a literature review and research opportunities},
  author={Quanbo Zha and Gang Kou and Hengjie Zhang and Haiming Liang and Xia Chen and Congcong Li and Yucheng Dong},
  journal={Financial Innovation},
  year={2020},
  volume={6},
  pages={1-22},
  url={https://api.semanticscholar.org/CorpusID:232042137}
}

@article{BA-FA-LNE:23,
  author={Bizyaeva, Anastasia and Franci, Alessio and Leonard, Naomi Ehrich},
  journal={IEEE Transactions on Automatic Control}, 
  title={Nonlinear Opinion Dynamics With Tunable Sensitivity}, 
  year={2023},
  volume={68},
  number={3},
  pages={1415-1430},
  doi={10.1109/TAC.2022.3159527}}

@article{NA-LJ:17,
  author={Nedić, Angelia and Liu, Ji},
  journal={IEEE Transactions on Automatic Control}, 
  title={On Convergence Rate of Weighted-Averaging Dynamics for Consensus Problems}, 
  year={2017},
  volume={62},
  number={2},
  pages={766-781},
  doi={10.1109/TAC.2016.2572004}}

@article{FA-GD:04,
title = {The role of network topology on extremism propagation with the relative agreement opinion dynamics},
journal = {Physica A: Statistical Mechanics and its Applications},
volume = {343},
pages = {725-738},
year = {2004},
issn = {0378-4371},
doi = {https://doi.org/10.1016/j.physa.2004.06.102},
url = {https://www.sciencedirect.com/science/article/pii/S0378437104008581},
author = {Frédéric Amblard and Guillaume Deffuant}
}

@article{CB-CA-AP-FV:24,
title = {Bounded confidence opinion dynamics: A survey},
journal = {Automatica},
volume = {159},
pages = {111302},
year = {2024},
issn = {0005-1098},
doi = {https://doi.org/10.1016/j.automatica.2023.111302},
author = {Carmela Bernardo and Claudio Altafini and Anton Proskurnikov and Francesco Vasca}
}

@article{ML-HD:19,
title = {Impact of temporal network structures on the speed of consensus formation in opinion dynamics},
journal = {Physica A: Statistical Mechanics and its Applications},
volume = {523},
pages = {1355-1370},
year = {2019},
issn = {0378-4371},
doi = {https://doi.org/10.1016/j.physa.2019.04.206},
author = {Mingwu Li and Harry Dankowicz}
}

@article{ZL-FB-MM:05,
  author={Zhiyun Lin and Francis, B. and Maggiore, M.},
  journal={IEEE Transactions on Automatic Control}, 
  title={Necessary and sufficient graphical conditions for formation control of unicycles}, 
  year={2005},
  volume={50},
  number={1},
  pages={121-127}
}

@ARTICLE{RZ-ZL-GC-WM:24,
  author={Zhang, Ruichang and Liu, Zhixin and Chen, Ge and Mei, Wenjun},
  journal={IEEE Transactions on Automatic Control}, 
  title={Convergence Analysis of Weighted-Median Opinion Dynamics With Prejudice}, 
  year={2025},
  volume={70},
  number={6},
  pages={4155-4162},
  doi={10.1109/TAC.2025.3530877}}

@article{LM-PGH-MAP:24,
title={A Weighted-Median Model of Opinion Dynamics on Networks},author={Lasse Mohr and Poul G Hjorth and M. A. Porter},journal={ArXiv},year={2024},volume={abs/2406.17552},doi={10.48550/arXiv.2406.17552}}

@article{BDOA-MY:19,
  title={Recent advances in the modelling and analysis of opinion dynamics on influence networks},
  author={B. D. O. Anderson and M. Ye},
  journal={International Journal of Automation and Computing},
  volume={16},
  number={2},
  pages={129--149},
  year={2019},
  publisher={Springer}
}

@Article{PJ-AM-NEF-FB:13d,
  author =	 {P. Jia and A. MirTabatabaei and N. E. Friedkin and
                  F. Bullo},
  title =	 {Opinion Dynamics and The Evolution of Social Power in
                  Influence Networks},
  journal =	 sirev,
  year =	 2015,
  volume =	 57,
  number =	 3,
  pages =	 {367-397},
  doi =		 {10.1137/130913250},
  funding =	 {ARO-ICB-W911NF-09-D-0001}
}

@Book{FB:21,
  author =	 {F. Bullo},
  title =	 {Lectures on Network Systems},
  month =	 sep,
  year =	 2021,
  edition =	 {{1.5}},
  publisher =	 {Kindle Direct Publishing},
  fbnote =	 {With contributions by J. Cort{\'e}s, F. D\"orfler, and
                  S. Mart{\'\i}nez},
  funding =	 {ARO-W911NF-11-1-0092, ARO-MURI-W911NF-15-1-0577,
                  AFOSR-FA9550-15-1-0138, CPS-1035917, CPS-1135819,
                  HDTRA1-19-1-0017},
  url =		 {http://motion.me.ucsb.edu/book-lns},
  ISBN =	 {978-1986425643},
}

@article{AJ-AS-ATS:10,
  author =	 {A. Jadbabaie and A. Sandroni and A. Tahbaz-Salehi},
  title =	 {Non-{Bayesian} Social Learning},
  journal =	 {Games and Economic Behavior},
  volume =	 76,
  number =	 1,
  pages =	 {210--225},
  year =	 2012,
  doi =		 {10.1016/j.geb.2012.06.001},
}

@article{AVP-RT:17,
  title =	 {A Tutorial on Modeling and Analysis of Dynamic Social
                  Networks. {Part I}},
  author =	 {A. V. Proskurnikov and R. Tempo},
  journal =	 {Annual Reviews in Control},
  volume =	 43,
  pages =	 {65-79},
  doi =		 {10.1016/j.arcontrol.2017.03.002},
  year =	 2017
}

@article{AVP-RT:18,
  title =	 {A Tutorial on Modeling and Analysis of Dynamic Social
                  Networks. {Part II}},
  author =	 {A. V. Proskurnikov and R. Tempo},
  journal =	 {Annual Reviews in Control},
  volume =	 45,
  pages =	 {166-190},
  doi =		 {10.1016/j.arcontrol.2018.03.005},
  year =	 2018
}

@article{CA:13,
  title =	 {Consensus problems on networks with antagonistic
                  interactions},
  author =	 {C. Altafini},
  journal =	 tac,
  volume =	 58,
  number =	 4,
  pages =	 {935--946},
  year =	 2013,
  doi =		 {10.1109/TAC.2012.2224251},
}

@Article{DA-AO:11,
  author =	 {D. Acemoglu and A. Ozdaglar},
  title =	 {Opinion Dynamics and Learning in Social Networks},
  journal =	 {Dynamic Games and Applications},
  year =	 2011,
  volume =	 1,
  number =	 1,
  pages =	 {3-49},
  doi =		 {10.1007/s13235-010-0004-1},
}

@article{JRPF:56,
  Author =	 {J. R. P. {French~Jr.}},
  Title =	 {A formal theory of social power},
  Journal =	 {Psychological Review},
  Pages =	 {181--194},
  Volume =	 63,
  Number =	 3,
  year =	 1956,
  fbnote =	 "was JF:56"
}

@Article{MHDG:74,
  author =	 {M. H. DeGroot},
  title =	 {Reaching a Consensus},
  journal =	 {Journal of the American Statistical Association},
  year =	 1974,
  volume =	 69,
  number =	 345,
  pages =	 {118-121},
  doi =		 {10.1080/01621459.1974.10480137}
}

@article{NEF-AVP-RT-SEP:16,
  author =	 {N. E. Friedkin and A. V. Proskurnikov and R. Tempo and
                  S. E. Parsegov},
  title =	 {Network Science on Belief System Dynamics Under Logic
                  Constraints},
  volume =	 354,
  number =	 6310,
  pages =	 {321--326},
  year =	 2016,
  journal =	 science,
  funding =	 {ARO-MURI-W911NF-15-1-0577},
  doi =		 {10.1126/science.aag2624},
}

@article{NEF-ECJ:90,
  title =	 {Social influence and opinions},
  author =	 {N. E. Friedkin and E. C. Johnsen},
  journal =	 {Journal of Mathematical Sociology},
  volume =	 15,
  number =	 {3-4},
  pages =	 {193--206},
  year =	 1990,
  doi =		 {10.1080/0022250X.1990.9990069}
}

@Article{RH-UK:02,
  author =	 {R. Hegselmann and U. Krause},
  title =	 {Opinion dynamics and bounded confidence models, analysis,
                  and simulations},
  journal =	 {Journal of Artificial Societies and Social Simulation},
  year =	 2002,
  volume =	 5,
  number =	 3,
  url =		 {http://jasss.soc.surrey.ac.uk/5/3/2.html},
}

@ARTICLE{SEP-AVP-RT-NEF:17,
  author =	 {S. E. Parsegov and A. V. Proskurnikov and R. Tempo and
                  N. E. Friedkin},
  journal =	 tac,
  title =	 {Novel Multidimensional Models of Opinion Dynamics in
                  Social Networks},
  year =	 2017,
  volume =	 62,
  number =	 5,
  pages =	 {2270-2285},
  doi =		 {10.1109/TAC.2016.2613905},
  funding =	 {ARO-MURI-W911NF-15-1-0577},
}

@Article{WR-RWB:05,
  author =	 {W. Ren and R. W. Beard},
  title =	 {Consensus seeking in multiagent systems under dynamically
                  changing interaction topologies},
  journal =	 tac,
  year =	 2005,
  volume =	 50,
  number =	 5,
  pages =	 "655-661",
  doi =		 {10.1109/TAC.2005.846556},
}

@article{CVK-SM-PG-PJR-JMH-VDB:16,
  author =	 {C. {Vande~Kerckhove} and S. Martin and P. Gend and
                  P. J. Rentfrow and J. M. Hendrickx and V. D. Blondel},
  fullauthor =	 {Vande Kerckhove, Corentin and Martin, Samuel and Gend,
                  Pascal and Rentfrow, Peter J. and Hendrickx, Julien
                  M. and Blondel, Vincent D.},
  journal =	 plosone,
  title =	 {Modelling Influence and Opinion Evolution in Online
                  Collective Behaviour},
  year =	 2016,
  month =	 06,
  volume =	 11,
  pages =	 {1-25},
  number =	 6,
  doi =		 {10.1371/journal.pone.0157685}
}

@article{MT:68,
  title={Towards a mathematical theory of influence and attitude change},
  author={M. Taylor},
  journal={Human Relations},
  volume={21},
  number={2},
  pages={121--139},
  year={1968},
  publisher={Sage Publications Sage CA: Thousand Oaks, CA}
}

@article{AN-AO:14,
  title={Distributed optimization over time-varying directed graphs},
  author={A. Nedi{\'c} and A. Olshevsky},
  journal={IEEE Transactions on Automatic Control},
  volume={60},
  number={3},
  pages={601--615},
  year={2014},
  publisher={IEEE}
}

@article{WM-FB-GC-JH-FD:22,
  title =	 {Micro-Foundation of Opinion Dynamics: Rich Consequences of an Inconspicuous       
             Change},
  author =	 {W. Mei and F. Bullo and G. Chen and J. Hendrickx and F. D\"{o}rfler},
  journal =	 {Physical Review Research},
  volume={4},
  number={2},
  pages={023213},
  year={2022},
  publisher={APS}
}

@article{WM-JMH-GC-FB-FD:24,
  title={Convergence, Consensus and Dissensus in the Weighted-Median Opinion Dynamics},
  author={W. Mei and J. M. Hendrickx and G. Chen and F. Bullo and F. D{\"o}rfler},
  journal={IEEE Transactions on Automatic Control},
  year={2024},
  publisher={IEEE}
}

@STRING{tac  = "IEEE Transactions on Automatic Control"}

@STRING{sirev = "SIAM Review"}

@STRING{automatica = "Automatica"}

@string{science = "Science"}

@STRING{springer = "Springer"}

@string{plosone = "PLoS One"}

@article{blanchini1999set,
  title={Set invariance in control},
  author={Blanchini, Franco},
  journal={Automatica},
  volume={35},
  number={11},
  pages={1747--1767},
  year={1999},
  publisher={Elsevier}
}

@article{redheffer1972theorems,
  title={The theorems of Bony and Brezis on flow-invariant sets},
  author={Redheffer, RM},
  journal={The American Mathematical Monthly},
  volume={79},
  number={7},
  pages={740--747},
  year={1972},
  publisher={Taylor \& Francis}
}

      % and a bib file to produce the 
                                 % bibliography (preferred). The
                                 % correct style is generated by
                                 % Elsevier at the time of printing.

%\begin{thebibliography}{99}     % Otherwise use the  
                                 % thebibliography environment.
                                 % Insert the full references here.
                                 % See a recent issue of Automatica 
                                 % for the style.
%  \bibitem[Heritage, 1992]{Heritage:92}
%     (1992) {\it The American Heritage. 
%     Dictionary of the American Language.}
%     Houghton Mifflin Company.
%  \bibitem[Able, 1956]{Abl:56}
%     B.~C.~Able (1956). Nucleic acid content of macroscope. 
%     {\it Nature 2}, 7--9. 
%  \bibitem[Able {\em et al.}, 1954]{AbTaRu:54}   
%     B.~C. Able, R.~A. Tagg, and M.~Rush (1954).
%     Enzyme-catalyzed cellular transanimations.
%     In A.~F.~Round, editor, 
%     {\it Advances in Enzymology Vol. 2} (125--247). 
%     New York, Academic Press.
%  \bibitem[R.~Keohane, 1958]{Keo:58}
%     R.~Keohane (1958).
%     {\it Power and Interdependence: 
%     World Politics in Transition.}
%     Boston, Little, Brown \& Co.
%  \bibitem[Powers, 1985]{Pow:85}
%     T.~Powers (1985).
%     Is there a way out?
%     {\it Harpers, June 1985}, 35--47.

%\end{thebibliography}

\end{document}